\documentclass[a4paper,fleqn,usenatbib]{mnras}

\usepackage[T1]{fontenc}
\usepackage{ae,aecompl}
\usepackage{graphicx}
\usepackage{amsmath}
\usepackage{amssymb}

\title[New insight into the MF of GCs]{New Insight into the stellar mass function of Galactic globular clusters}

\author[Ebrahimi et al.]
{H. Ebrahimi$^{1}$\thanks{E-mail:  \mbox{h.ebrahimi@iasbs.ac.ir} (HE)},
A. Sollima$^{2}$,  
H. Haghi$^{1}$, 
H. Baumgardt$^{3}$, 
M. Hilker$^{4}$ \\
$^{1}$Department of Physics, Institute for Advanced Studies in Basic Sciences (IASBS), PO Box 11365-9161, Zanjan, Iran\\
$^{2}$INAF Osservatorio Astrofisica e Scienza dello Spazio, via Gobetti 93/3, I-Bologna 40129, Italy \\
$^{3}$School of Mathematics and Physics, The University of Queensland, St. Lucia, QLD 4072, Australia \\
$^{4}$European Southern Observatory, Karl-Schwarzschild-Str. 2, D-85748 Garching, Germany }
\begin{document}

\date{Accepted 2020 March 31. Received 2020 February 9; in original form 2019 October 24}

\pagerange{\pageref{firstpage}--\pageref{lastpage}} \pubyear{2017}

\maketitle

\label{firstpage}

\maketitle

\begin{abstract}

We present the results of the analysis of deep photometric data of 32 Galactic
globular clusters. We analysed 69 parallel field images 
observed with the Wide Field Channel of the Advanced Camera for Surveys of 
the Hubble Space Telescope which complemented the already available photometry 
from the globular cluster treasury project covering the central regions of 
these clusters.
This unprecedented data set has been used to calculate
the relative fraction of stars at different masses (i.e. the present-day mass
function) in these clusters by comparing the observed distribution of stars
along the cluster main sequence and across the analysed field of view with
the prediction of multimass dynamical models.
For a 
subsample of 31 clusters, we were able to obtain also the  
half-mass radii, mass-to-light ratios and the mass fraction of dark 
remnants using available radial velocity information. We found that the 
majority of globular clusters have single power law mass functions 
$F(m)\propto m^\alpha$ with slopes $\alpha>-1$ 
in the mass range $0.2<m/\text{M}_{\odot}<0.8$. 
By exploring the correlations between the 
structural/dynamical 
and orbital parameters, 
we confirm the tight anticorrelation between the mass function slopes and 
the half-mass relaxation times already reported in previous works, and 
possible second-order dependence on the cluster metallicity. This might 
indicate the relative importance of both initial conditions and evolutionary 
effects on the present-day shape of the mass function.

\end{abstract}

\begin{keywords}
methods: numerical - techniques: photometric - techniques: radial velocities -
stars: kinematics and dynamics - stars: luminosity function, mass function - globular clusters:
general
\end{keywords}

\section{Introduction} \label{Sec:1}
Globular clusters (GCs) are among the most useful objects to study stellar 
astrophysics. Consisting of a large number of stars with similar ages and 
chemical compositions they are valuable to test stellar evolution models. 
Beside their importance in stellar evolution studies, GCs are the oldest 
collisional systems such that dynamical drivers (like two-body relaxation) 
and external tidal field have changed their 
structures. By comparing observational coordinates in phase-space (projected 
positions and three-dimensional motions) with theoretical 
models, one can derive many {\textbf important} parameters of GCs like their masses, 
mass functions and degree of mass segregation.

The present-day structure of GCs depends directly on the mass distribution of stars 
that extends from the faintest stars at the hydrogen burning limit (with 
masses around 0.1 M$_{\odot}$) to massive black holes (with masses larger than 
15 M$_\odot$). The shape of the present-day mass function (MF) is the result of 
the effect of complex mechanisms of dynamical and stellar evolution from the initial mass function (IMF). 
The universality of the IMF is a highly debated topics in astrophysics 
\citep{bastian2010, kroupa2013}. The first attempt to introduce a parameterized IMF as a 
single power-law function was made by \cite{salpeter1955} and followed by 
a log-normal 
\citep{miller1979, chabrier2003}, a multi-segment power-law 
\citep{kroupa1993, kroupa2001} and a tapered power-law IMF \citep{demarchi2005}. 
In recent decades, many studies have tried to fit the above functions to the
observed MF of various unevolved groups of stars, e.g. field stars \citep{czekaj2014,rybizki2015,mor2019,sollima2019}, young and 
embedded clusters \citep{weights2009,weisz2013}, OB associations \citep{massey2003,dario2012}, open 
clusters \citep{moraux2003,sheikhi2016}, and dwarf
galaxies \citep{cappellari2006,gennaro2018}. Most studies of resolved stellar
populations in the disk of the Milky Way showed that stars form following an 
IMF that has a universal form \citep{kroupa2001,kroupa2002} which is referred 
to as the "canonical" IMF. This poses a problem for star formation theories 
which predict a dependence on the environment where star formation takes place \citep{kroupa2013,chabrier2014}. 
In spite of no evidence of significant variations 
of the IMF in these studies, the universality of the IMF is still a matter of debate.
In particular, recent integrated light spectroscopic studies in the centres of giant 
ellipticals seem to favour a bottom-heavy IMF \citep{conroy2017, vandokkum2017}, 
while some extragalactic super starburst regions indicate a top-heavy IMF \citep{zhang2018}. 

From the IMF to the present-day MF, the evolution of GC MFs depends on 
many processes. At early stages, stellar evolution leads to the disruption of the 
most massive stars depriving the high-mass tail of the IMF. 
On long timescale, gravitational encounters among the stars and the interaction of the GC with 
the external tidal field also alter the shape of the MF. 
The tendency toward kinetic energy equipartition leads high-mass 
stars to sink into the core and low-mass stars to migrate toward the outskirts 
\citep{spitzer1987}. 
Low-mass stars can gain enough energy to escape from the GC more 
efficiently than high-mass ones, with a consequent depletion of the low-mass 
end of the MF \citep{lamers2013}. As a result, as the 
GC loses a significant fraction of its low-mass stars, the faint end of its IMF evolves from its initial 
toward a flatter shape \citep{baumgardt2003}. 
This process is enhanced by the tidal force exerted by the host galaxy which 
accelerates the process of mass loss \citep{gieles2011}.
This so-called "dynamical" mass segregation might be accompained by 
a "primordial" mass segregation. The evidence and effect of this latter kind of mass 
segregation has been widely discussed on the basis of both observational 
\citep{frank2014} and theoretical grounds
\citep{haghi2014,haghi2015}. For instance, \cite{zonoozi2011, zonoozi2014, zonoozi2017} have shown that 
to explain the MF flattening 
in the centres of Pal 4 and Pal 14, the presence of both types of mass segregation is 
necessary. The effect of the various drivers on the evolution of GC MFs 
have been considered by many studies including many further ingredients, e.g. 
Galactic disc/bulge shocking \citep{ostriker1972,aguilar1988}, binary star evolution and 
interaction \citep{schneider2015} and orbital eccentricity \citep{madrid2014,webb2014}.

\begin{figure*}
\begin{center}
\includegraphics[width=180mm]{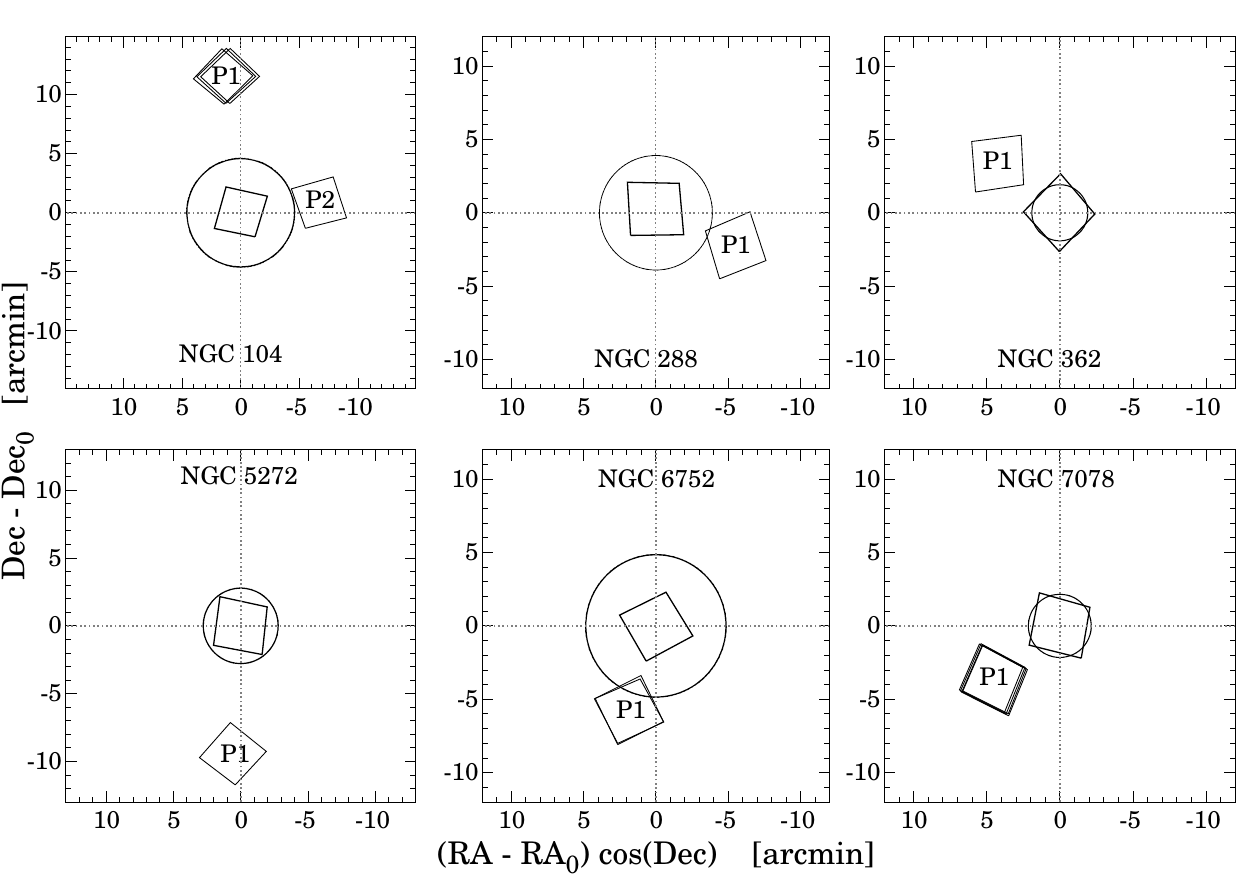}
\caption{Maps of the pointings around the 6 GCs analysed here and not present in the 
\protect\cite{simioni2018} sample. 
The squares in the centre show the regions covered by the HST Treasury program. The 
squares labelled by 'P' mark the location of the parallel pointings. The half-mass radii (from the present work) are marked with circles. 
The maps of the other GCs of our sample are available in \protect\cite{simioni2018}.}
\label{map}
\end{center}
\end{figure*}

The proper modelling of mass segregation as a function of radius is crucial to estimate the 
global MF of a star cluster. Indeed, observations are often localized in a 
restricted portion of the GC, and suitable corrections are necessary to 
account for such MF variations \citep{paust2010}. 

From a theoretical perspective, there are a few methods to investigate the 
dynamical evolution of gravitional systems: N-body simulations \citep{baumgardt2003,webb2015}, Fokker-Planck models 
\citep{takahashi2000,murphy2011}, Monte Carlo models \citep{giersz2001,joshi2001} and through analytic 
methods \citep[e.g. like the \textsc{emacss} code][]{alexander2014}. All these models have been used to
study the mechanisms driving the evolution from the IMF to the present-day MF. 

During the last decades, the \emph{Hubble Space Telescope} (HST) 
has allowed us an unprecedented insight into the GC stellar 
populations.  
Within the \emph{HST Treasury program}, 
\citet{sarajedini2007} performed uniform photometry of stars in the central region 
of 65 GCs sampling stars as faint as 0.2 M$_{\odot}$ with a 
$S/N\gtrsim$ 10. This program has served many studies aimed at testing 
stellar evolution, isochrones, luminosity functions and 
synthetic horizontal-branch models \citep{dotter2007}, also deriving the ages of 
$\sim$60 GCs \citep{marinfranch2009,dotter2010} and the MF of a sample of 17 
GCs \citep{paust2010}. 
Moreover, the 
\emph{HST UV legacy survey of Galactic GCs program} designed to find multiple 
stellar populations extended the observational datasets to ultraviolet wavelengths  
\citep{piotto2015, milone2017}. In the context of this 
project, \cite{simioni2018} presented a photometric catalogues of 110 
parallel fields in the outskirts of 48 Galactic GCs. By combining 
the catalogues of the two above mentioned programs, the photometric catalogues 
of 56 Galactic GCs have been made public \citep{nardiello2018}. The second data 
release (DR2) of the \emph{Gaia} mission has also improved the observational 
scenario of GCs. Among the recent works based on \emph{Gaia} DR2, 
\cite{baumgardt2019} have derived the mean proper motions and space velocities 
of 154 Galactic GCs and the velocity dispersion profiles of 141 GCs. 

\begin{table}
    \centering
    \caption{Observing logs.}
    \label{table1}
    \begin{tabular}{cccccccccc}
        \hline 
        \hline     
        NGC & Pointing & Filter & Exposure time (s)  \\
        \hline  
        \hline
        104 & 1 & F475W & 1050; 2$\times$986; 947; 2$\times$880  \\
        & & F814W & 877; 870; 806; 800; 767; 760  \\
         & 2 & F606W & 1498; 1457; 1443  \\
          &  &  & 1442; 1385; 1371  \\
        & & F814W & 1457;1371; 1358; 1357  \\
        & &  & 1303; 1118; 100  \\
        \hline    
        288 & 1 & F606W & 3$\times$200; 15  \\
        & & F814W & 3$\times$150; 10  \\
        \hline
        362 & 1 & F606W & 172; 86; 39  \\
        & & F814W & 60; 25  \\
        \hline
        1261 & 1 & F475W & 770  \\
        & & F814W & 694  \\
         & 2 & F475W & 745  \\
        & & F814W & 669  \\
         & 3 & F475W & 766  \\
        & & F814W & 690  \\
         & 4 & F475W & 745  \\
        & & F814W & 669  \\
         & 5 & F475W & 829  \\
        & & F814W & 753  \\
        \hline
        1851 & 1 & F475W & 2$\times$1277; 1237; 2$\times$40  \\
        & & F814W & 6$\times$488; 40  \\
         & 2 & F475W & 2$\times$1277; 2$\times$1237; 2$\times$40  \\
        & & F814W & 8$\times$488; 2$\times$40  \\
        \hline
        2298 & 1 & F475W & 2$\times$785  \\
        & & F814W & 2$\times$683  \\
         & 2 & F475W & 887; 885  \\
        & & F814W & 816; 815  \\
        \hline 
        3201 & 1 & F475W & 685  \\
        & & F814W & 612  \\
         & 2 & F475W & 689  \\
        & & F814W & 616  \\
        \hline
        4590 & 1 & F475W & 627  \\
        & & F814W & 554  \\
         & 2 & F475W & 627  \\
        & & F814W & 554  \\
        \hline
        5024 & 1 & F475W & 4$\times$725; 2$\times$723  \\
        & & F814W & 3$\times$370  \\
         & 2 & F475W & 4$\times$775; 2$\times$774  \\
        & & F814W & 3$\times$375  \\
        \hline 
        5053 & 1 & F475W & 740  \\
        & & F814W & 664  \\
         & 2 & F475W & 740  \\
        & & F814W & 664  \\
         & 3 & F475W & 790  \\
        & & F814W & 714  \\
         & 4 & F475W & 790  \\
        & & F814W & 714  \\
         & 5 & F475W & 765  \\
        & & F814W & 689  \\
        \hline
        5272 & 1 & F475W & 3$\times$800  \\
        & & F814W & 3$\times$760  \\
        \hline  
        5286 & 1 & F475W & 728  \\
        & & F814W & 655  \\
         & 2 & F475W & 603  \\
        & & F814W & 559  \\ 
        \hline
        \hline   
    \end{tabular}   
\end{table}
\begin{table}
    \centering
    \contcaption{}
    \begin{tabular}{cccccccccc}
        \hline 
        \hline     
        NGC & Pointing & Filter & Exposure time (s)  \\
        \hline  
        \hline
        5466 & 1 & F475W & 835; 834  \\
        & & F814W & 765; 763  \\
         & 2 & F475W & 2$\times$776  \\
        & & F814W & 2$\times$700  \\
        \hline  
        5897 & 1 & F475W & 833; 830  \\
        & & F814W & 2$\times$761  \\
         & 2 & F475W & 781; 779  \\
        & & F814W & 710; 709  \\
        \hline  
        5904 & 1 & F475W & 620  \\
        & & F814W & 559  \\
         & 2 & F475W & 621  \\
        & & F814W & 559  \\
        \hline  
        5986 & 1 & F475W & 676  \\
        & & F814W & 603  \\
         & 2 & F475W & 2$\times$603  \\
        & & F814W & 2$\times$559  \\
        \hline  
        6093 & 1 & F475W & 5$\times$845; 5$\times$760   \\
        & & F814W & 5$\times$539  \\
        \hline  
        6101 & 1 & F475W & 762  \\
        & & F814W & 686  \\
         & 2 & F475W & 762  \\
        & & F814W &  686  \\
         & 3 & F475W & 800  \\
        & & F814W & 724  \\
         & 4 & F475W & 851  \\
        & & F814W &  775  \\
        & 5 & F475W & 800  \\
        & & F814W &  724  \\
        \hline  
        6144 & 1 & F475W & 679  \\
        & & F814W & 606  \\
         & 2 & F475W & 679  \\
        & & F814W &  606  \\
        \hline  
        6218 & 1 & F475W & 721  \\
        & & F814W & 648  \\
         & 2 & F475W & 645  \\
        & & F814W &  572  \\
        \hline  
        6254 & 1 & F475W & 721  \\
        & & F814W & 648  \\
         & 2 & F475W & 644  \\
        & & F814W &  571  \\
        \hline  
        6341 & 1 & F475W & 638  \\
        & & F814W & 565  \\
         & 2 & F475W & 750  \\
        & & F814W &  677  \\
         \hline  
        6362 & 1 & F475W & 651  \\
        & & F814W & 578  \\
         & 2 & F475W & 760  \\
        & & F814W &  687  \\
         \hline  
        6541 & 1 & F475W & 689  \\
        & & F814W & 616  \\
         & 2 & F475W & 639  \\
        & & F814W &  566  \\
        \hline  
        6584 & 1 & F475W & 640  \\
        & & F814W & 567  \\
         & 2 & F475W & 726  \\
        & & F814W &  653  \\
        \hline
        \hline   
    \end{tabular}   
\end{table}
\begin{table}
    \centering
    \contcaption{}
    \begin{tabular}{cccccccccc}
        \hline 
        \hline     
        NGC & Pointing & Filter & Exposure time (s)  \\
        \hline  
        \hline
        6723 & 1 & F475W & 666  \\
        & & F814W & 592  \\
         & 2 & F475W & 626  \\
        & & F814W &  551  \\
        & 3 & F475W & 624  \\
        & & F814W &  551  \\
        \hline
        6752 & 1 & F606W & 500  \\
        & & F814W & 200; 15   \\
        \hline
        6779 & 1 & F475W & 731  \\
        & & F814W & 658  \\
         & 2 & F475W & 637  \\
        & & F814W &  564  \\
        \hline
        6809 & 1 & F475W & 753  \\
        & & F814W & 680  \\
         & 2 & F475W & 677  \\
        & & F814W &  604  \\
        \hline
        7078 & 1 & F475W & 4$\times$702  \\
        & & F814W & 4$\times$121   \\
        \hline
        7089 & 1 & F475W & 717  \\
        & & F814W & 643  \\
         & 2 & F475W & 668  \\
        & & F814W &  593  \\
        & 3 & F475W & 611  \\
        & & F814W &  534  \\
        \hline
        7099 & 1 & F475W & 656  \\
        & & F814W & 583  \\
         & 2 & F475W & 656  \\
        & & F814W &  583  \\
        \hline  
        \hline                
    \end{tabular}   
\end{table}     

By comparing observational data and simple dynamical models, 
several studies have attempted to calculate GC MFs and evaluate possible 
correlations between GC parameters. Among the most comprehensive studies, 
but based on heteregeneous measurements, 
\citet{capaccioli1993} and \citet{djorgovski1993} reported 
a dependence of the MF slopes 
on the Galactocentric distances and on the heights above the Galactic plane. 
\citet{piotto1999} analysed deep HST images taken near the half-mass 
radii of seven GCs and found that the MF slopes correlate with the orbital 
destruction rates of the clusters and anticorrelate with their half-mass 
relaxation times, although their small sample hampered any firm conclusion on 
the significance of these correlations. 
\citet{demarchi2007}, derived the MF of 20 GCs using HST and 
Very Large 
Telescope (VLT) data, reporting a well-defined correlation between the slope of 
their MFs and their King model concentration parameter c. 
\cite{paust2010} 
 provided the MF for 17 GCs comparing the luminosity function derived from the 
 \emph{HST Treasury program} data and multimass models and found 
 that the MF slope correlates with central density, but with neither metallicity nor 
 Galactic location. Using to the same data set and a similar technique, Sollima \& Baumgardt 
 (\citeyear{sollima2017b}; hereafter \citeauthor{sollima2017b}) expanded 
 the sample to 35 GCs. They determined the structural and dynamical parameters 
 of 29 GCs with available radial velocity information and revealed a tight 
 anticorrelation between MF slopes and half-mass relaxation times and correlation with the
 dark remnant fractions. They concluded that the internal dynamical evolution 
 is the main responsible in shaping the present-day MFs. These last works, 
 while representing the most complete census of MF based on the 
 deepest photometric data set available so far, suffer from the lack of 
 information outside the GC cores and constrain the model predictions only 
 with the MF measured in the cluster centre. This can potentially lead to 
 significant bias in the estimated MF, in particular in those GCs whose 
 extent exceed the field of view of the available data \citep{sollima2015}.
 By fitting a large set of N-body simulations to their velocity dispersion 
 and surface density profiles, the correlation between MF slope and 
 relaxation time has been confirmed by Baumgardt \& Hilker
 (\citeyear{baumgardt2018}; hereafter \citeauthor{baumgardt2018}).          

In this paper, we present the results of the photometric analysis of 69 
HST/ACS/WFC parallel fields for 32 Galactic GCs and combine them with the 
available data from the \emph{HST Treasury program} in the central regions of these 
GCs. This unprecedented data set is used to derive the global present-day MFs 
through the comparison with multimass analytical models. 
 
This paper is organized as follows: in section \ref{Sec:2} the observational 
material is presented and the data reduction technique is described. 
The multimass dynamical model, the algorithm of MF determination and fitting 
technique are described in section \ref{Sec:3} and \ref{Sec:4}. In section \ref{Sec:5} we 
present the derived GCs MF and look for the 
possible correlations with various parameters. We summarize our results in 
Section \ref{Sec:6}.         
\section{Photometry and data reduction}\label{Sec:2}  

Images have been obtained 
with the Advanced Camera for Surveys (ACS) on board HST using the single 
channel camera WFC. The detector includes two similar chips of 2048 $\times$ 4096 pixels each with 
a pixel-scale of 0.05 arcsec per pixel. Therefore the entire channel covers 
an effective field of view of 202 arcsec $\times$ 202 arcsec. 
The photometric catalogues in the
F606W and F814W filters of the region around the centers of 65 GCs have been 
published as part of the 
\emph{HST Treasury program} \citep{sarajedini2007}. The results of artificial star 
experiment performed on this data set have been also provided by \citet{anderson2008}.

\begin{figure*}
\begin{center}
\includegraphics[width=180mm]{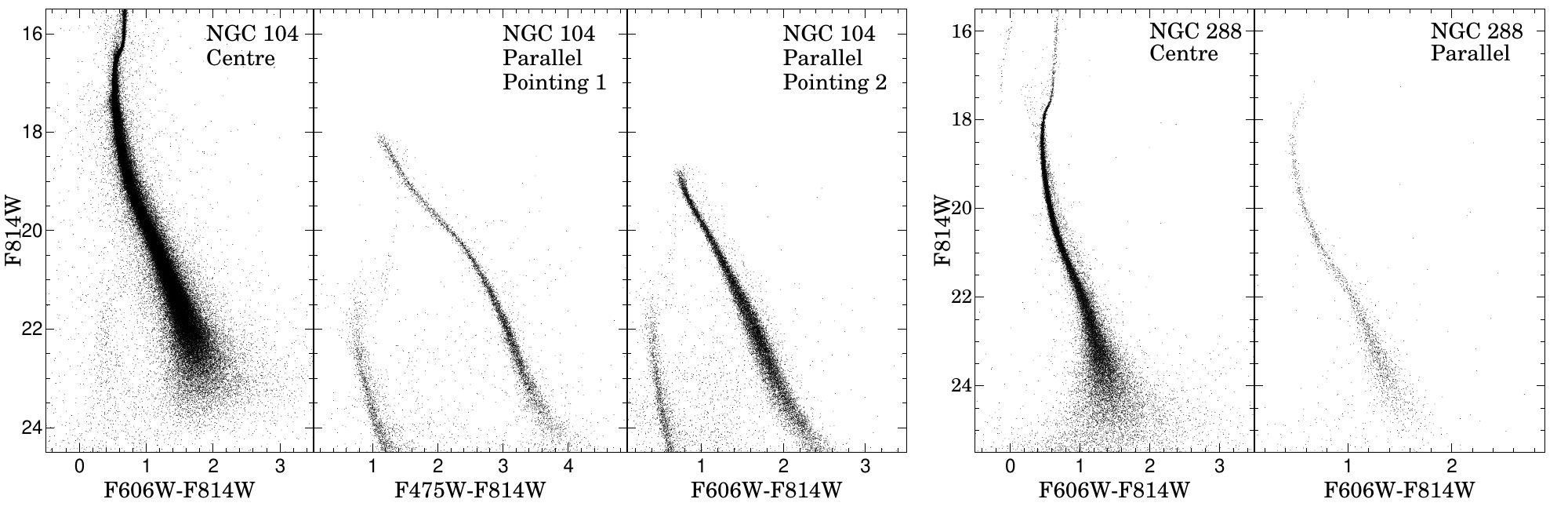}
\caption{CMD of NGC 104 (left panels) NGC 288 (right panels). The same plot for the entire sample of 
GCs is available as supplementary material in the online version of the paper.
}
\label{cmd1}
\end{center}
\end{figure*}

Parallel pointings for all the GCs included in the \emph{HST Treasury project} are also 
available as ancillary products of the \emph{HST UV legacy survey of Galactic GCs program}. 
In a recent work, \citet{simioni2018} provided accurate photometry 
for these pointings together with cluster membership probabilities.
Unfortunately, no artificial star experiments (essential for our purpose to 
estimate the catalog completeness at different magnitudes and in different 
crowding conditions) were performed by these authors.
So, we performed an independent photometric analysis of this data set and the 
artificial star experiments consistently.
Among the entire sample of GCs included in the \emph{HST Treasury project}, we excluded 
 NGC 4147 and NGC 6205 because of the lack of parallel field images available 
until 2018. After performing the photometric process, we 
exclude some further GCs because of (i) contamination by the Sagittarius stream (e.g. 
NGC 6652 and NGC 6681) or the Sagittarius dwarf galaxy (e.g. NGC 6715) or the bulge 
(e.g. NGC 6624 and NGC 6637), (ii) large helium variation ($\Delta Y>0.1$; e.g. NGC 2808, NGC 5139, NGC 6388 and 
NGC 6441) affecting the mass-luminosity relation along the main sequence, and (iii) 
a completeness level below 10\% at the hydrogen-burning limit 
(e.g. NGC 6171, NGC 6934 and NGC 6981). A final sample of 32 GCs passed the above criteria (see Table \ref{table1}). 
The CMDs of the central region and of the parallel fields of these GCs are shown in 
Fig. \ref{cmd1}\footnote{The CMDs of all the GCs are available as supplementary material in the online version of the paper.}. 

We retrieved images of 69 parallel fields (pointings) in the outer region 
of these 32 GCs 
provided by the HST/ACS/WFC public archive and released before 2018. These pointings 
are centered approximately from 6 arcmin to 12 arcmin from the center of the target 
GCs. In Fig \ref{map}, the maps of the region sampled by those observations are shown. 
Each pointing consists of several images differing from each other because of 
their different 
filters and exposure times. We selected them in order to ensure at least two images 
observed through two different filters (including F814W) and with an exposure time long 
enough to allow to construct a deep 
colour-magnitude diagram (CMD) sampling stars with masses down to $\sim 0.15$ M$_{\odot}$. 
In our collected samples, each parallel pointing has only one filter pair: 
(F606W, F814W) or (F475W, F814W). All GCs with multiple parallel pointings 
(with the exception of NGC 104) have been observed with the same filter pair. 
For simplicity, in spite of their different throughput, we will name the 
F475W/F606W magnitudes as V magnitude and the F814W magnitude as I 
magnitude. The logs of the observations are listed in Table \ref{table1}. 

We performed the photometric analysis on the flat-fielded HST/ACS/WFC images 
corrected for CTE (\texttt{flc}) while the drizzled 
(\texttt{drc}) images were used for source detection. 
We employed the
stellar photometry package \textsc{dolphot} {\small v2} optimized 
in the \texttt{HSTphot} extention \citep{dolphin2000} to analyse ACS 
images \citep{dolphin2016}. 
A high-signal to noise image has been created by aligning and 
stacking all available \texttt{drc} images of each pointing  
and used to create a list of detected 
sources. 
This reference image has then been used as input by 
the \textsc{dolphot} point spread function 
(PSF)-fitting routine which has been run on individual 
images. Aperture corrections have been computed using the 
most isolated and bright sources and applied to the output catalogue.
In this catalog, the average magnitude of each detected star and those measured in 
all the individual \texttt{flc} images were listed together with quality flags, S/N, 
roundness, sharpness and star positions in pixels. To select an optimal set of 
stars, we set 
 the following criteria: (i) at least two reasonable magnitudes should exist 
 per filter for each star in each pointing, (ii) an absolute value of sharpness 
 smaller than 0.1;\footnote{This criterion came from the definition of sharpness 
 \citep{dolphin2000}: the sharpness of completely flat, sharp and 
 perfectly-fit stars are -1, 1 and 0, respectively.} (iii) an object type flag 
 equal to 1, which corresponds to good stars; (iv) a photometry quality 
 flag less than 4, which includes stars with negligible photometric 
 errors. 
We cross-correlated our catalogues with those provided by \citet{simioni2018} to 
convert {\textbf ($x,~y$)} coordinates into the standard astrometric reference system {\textbf ($RA,~Dec$)}.                       
A star-by-star comparison between our photometry and that of 
\citet{simioni2018} indicates average differences between the two datasets 
of $\Delta V,\Delta I(\text{Present work - Simioni et al.}) = 0.04$ in 
V and I passbands for all the GCs of the sample, indicating a small shift 
in the calibration process.
   
To perform artificial star experiments for a pointing in each GC, an input set 
of fake stars has been simulated. The positions of fake stars have been chosen by 
defining a regular grid of $190\times 190$ cells separated by 20 pixels 
in each direction, which cover homogeneously the ACS/WFC field of view. 
Artificial stars have been placed at a random position within a $16\times 16$ pixel square centered around each grid knot
(one star per cell). Given the typical size of the ACS/WFC PSF ($FWHM\sim2~px$) 
this ensures that artificial stars cannot blend each other (self-crowding). 
The magnitudes of stars have been extracted from 
the theoretical isochrone provided by \cite{dotter2007} with 
the suitable age, metallicity and alpha-element abundance 
of each GC \citep{dotter2010}. The absolute magnitudes of stars have been converted into the 
apparent ones by adopting the distance moduli and 
reddening determined by \cite{dotter2010}. The masses of fake stars were extracted 
from a Kroupa IMF \citep{kroupa2001} covering the full range of masses 
covered by the isochrone from the hydrogen burning limit up to the red giant branch tip. 
The corresponding magnitudes of fake stars have been used by 
\textsc{dolphot}
to normalize the fluxes of 
the PSF which have been placed into the original science images and
the photometric analysis has then been performed on synthetic images with 
the same prescriptions adopted for the original frames.
With the above procedure, 36,100 fake stars have been 
simulated per run. Moreover, several (from 1 to 5, depending on the number of GC pointings) 
independent runs have been performed in each GC and the artificial star catalogues of 
different runs have been stacked together in a unique list containing from 
36,100 to 180,500 artificial stars.
An artificial star has been considered as recovered if the difference between its 
input and output magnitudes is smaller than $2.5 \log(2)\sim0.752$ mag in both 
V and I magnitudes.                

For a subsample of 31 GCs we were able to derive the dynamical parameters through the 
comparison with the set of available radial velocities. For this purpose we used 
the sample of radial velocities derived by \citeauthor{baumgardt2018}\footnote{https://people.smp.uq.edu.au/HolgerBaumgardt/globular/} 
which have been 
obtained from a combination of more than 45,000 high-resolution spectra observed with 
FLAMES@VLT and DEIMOS@Keck stars with literature data. For the 31 GCs in our sample, the number 
of stars with available radial velocities ranges from 19 (for the lowest mass GC 
NGC 6144) to 2867 (for the most massive GC NGC 104). 

\section{Models}\label{Sec:3}

To convert the relative distrubution of stars at different radii into the 
global MF, a suitable dynamical model is needed.
The structure of multimass stellar systems can be obtained from the
phase-space distribution of stars proposed by \cite{michie1963} and \cite{king1966}.
The isotropic form of this distribution is \citep{gunn1979}:
\begin{equation} \label{dist-eq}
f(E)=\sum\limits^H_{i=1}k_{i}f(m_{i},r,v)=\sum\limits^H_{i=1}k_{i}\bigg[\exp \bigg(\frac{-A_{i}E}{\sigma_{K}^{2}}\bigg)-1\bigg],
\end{equation}   
where $H$ is the number of mass components, $m_i$ is the mass of the $i$th component, 
$v$ and $r$ are the 3D velocity and distance from the GC centre, $k_i$ are 
coefficients determining the number of stars in each mass bin, $A_i$ are 
coefficients regulating the equilibrium among the stars with different masses 
and $\sigma_{K}^{2}$ is a constant determining the normalization in energy.
$E=v^{2}/2+\psi(r)$ is the energy per unit mass, where $\psi$ is the effective 
potential defined as the difference between the potential at an arbitrary radius and
the potential at the GC tidal radius, $r_t$. 
Following \citet{gunn1979}, we adopted $A_i\propto m_{i}$. 
A degree of 
radial anisotropy can be accounted for by multiplying the above distribution
function by a term depending on the angular momentum. 
\cite{watkins2015} analysed 
the HST proper motions of a sample of 22 nearby GCs and showed that 
deviations from isotropy are negligible
within GC cores and only a mild anisotropic velocity distributions near the 
half-mass radius is detectable. Thus, to limit the number of free parameters, we adopted the isotropic 
form of the distribution functions (eq. \ref{dist-eq}).

By integrating the distribution function over $v$ and $r$, the 3D number density, velocity
dispersion, number of stars, projected number density and line-of-sight velocity dispersion
can be derived for each mass group, respectively, as:
\begin{equation} 
n_i(r)=\int_0^{\sqrt{-2\psi(r)}} \, 4\pi v^{2}k_{i}f(m_{i},r,v)\,dv,
\end{equation}
\begin{equation}
\sigma_{v,i}^{2}(r)=\frac{1}{n_i(r)}\int_0^{\sqrt{-2\psi(r)}} \, 4\pi v^{4}k_{i}f(m_{i},r,v)\,dv,
\end{equation}
\begin{equation} \label{num-eq}
N_i=\int_0^{r_t} \, 4\pi r^{2}n_{i}(r)\,dr,
\end{equation}
\begin{equation} \label{surf-den-eq}
\Gamma(m_{i},R)=2\int_R^{r_t} \, \dfrac{rn_i(r)}{\sqrt{r^{2}-R^{2}}}\,dr,
\end{equation} 
\begin{equation}\label{los-vel-eq}
\sigma_{\text{LOS},i}^{2}(R)=\dfrac{2}{3\Gamma_{i}(R)}\int_R^{r_t} \, \dfrac{rn_{i}(r)\sigma_{v,i}^{2}(r)}{\sqrt{r^{2}-R^{2}}}\,dr,
\end{equation}
while the mean mass is given by
\begin{equation}
\bar{m}=\frac{\sum\limits^H_{i=1}N_i m_i}{\sum\limits^H_{i=1} N_i}
\label{m-eq}
\end{equation}   
The radial gradient of $\psi(r)$ is given by the Possion equation as:
\begin{equation}
\nabla^{2}\psi(r)=4\pi G\rho(r),
\end{equation} 
while the global 3D density is determined by the following expression:
\begin{equation}
\rho(r)=\sum\limits^H_{i=1}m_{i}n_{i}(r).
\end{equation}
To integrate the above equations, appropriate boundary conditions need to be set. 
At the center the value of the potential ($W_0 \equiv -\psi_0 /\sigma_k^{2}$) has been left as a free parameter while 
its derivative has been set to $d\psi/dr(0)=0$. 
At the tidal radius, defined as the distance where both density and potential vanish, 
we set $\psi(r_{t})=0$.

Along with $k_i$, two additional free parameters complete the definition of
the models: $r_c\equiv\sqrt{9\sigma_k^{2}/4\pi G\rho_0}$ and $\sigma_k^{2}$
which determine the size and mass of the system. 
The total mass, luminosity and surface brightness profile can be obtained respectively as:
\begin{equation}
M=\sum\limits^H_{i=1}N_{i}m_{i},
\end{equation}
\begin{equation} \label{lv-equ}
L_V=\sum\limits^H_{i=1}N_{i} {\mathcal{F}}_{i},
\end{equation}
\begin{equation}
\mu=-2.5 \log\Bigg( \sum\limits^H_{i=1} \Gamma_{i}{\mathcal{F}}_{i} \Bigg),
\end{equation}
where $\mathcal{F}_{i}$ is the average V-band flux of stars in the $i-$th component. 

\section{Method}  \label{Sec:4}

\begin{table*}
    \centering
    \caption{Parameters of the best-fitting models.}
    \label{table_main}
    \begin{tabular}{cccccccccc}  
        \hline     
        NGC & $\alpha$ & $\log(M_{\text{lum}}/{\text{M}}_{\odot})$ & $\log(M_{\text{dyn}}/{\text{M}}_{\odot})$ & $r_h$ & $f_{\text{remn}}$ & $t_{rh}$ & $M_{\text{dyn}}/L_V$ & $r_J$ \\
        & & & & (pc) & & (Gyr) & $({\text{M}}_{\odot}/\text{L}_{\odot})$ & (pc)\\
        \hline
        104 & -0.45 $\pm$ 0.12 & 5.574 $\pm$ 0.008 & 5.958 $\pm$ 0.012 & 6.024 $\pm$ 0.032 & 0.586 $\pm$ 0.014 & 4.946 & 2.022 $\pm$ 0.068 & 183.96 \\
        288 & -0.75 $\pm$ 0.05 & 4.638 $\pm$ 0.056 & 5.024 $\pm$ 0.033 & 10.121 $\pm$ 0.292 & 0.589 $\pm$ 0.061 & 5.075 & 2.084 $\pm$ 0.311 & 137.32 \\
        362 & -0.79 $\pm$ 0.05 & 5.110 $\pm$ 0.011 & 5.517 $\pm$ 0.029 & 4.811 $\pm$ 0.071 & 0.608 $\pm$ 0.028 & 2.455 & 1.339 $\pm$ 0.096 & 208.88 \\
        1261 & -0.72 $\pm$ 0.04 & 4.912 $\pm$ 0.011 & 5.260 $\pm$ 0.044 & 6.882 $\pm$ 0.064 & 0.551 $\pm$ 0.047 & 3.167 & 1.489 $\pm$ 0.156 & 190.57 \\
        1851 & -0.74 $\pm$ 0.04 & 5.113 $\pm$ 0.009 & 5.606 $\pm$ 0.026 & 5.631 $\pm$ 0.015 & 0.678 $\pm$ 0.020 & 3.375 & 1.861 $\pm$ 0.118 & 234.49 \\
        2298 & -0.05 $\pm$ 0.06 & 4.279 $\pm$ 0.011 & 4.596 $\pm$ 0.161 & 5.856 $\pm$ 0.324 & 0.518 $\pm$ 0.179 & 1.292 & 0.852 $\pm$ 0.317 & 98.74 \\
        3201 & -1.22 $\pm$ 0.10 & 4.936 $\pm$ 0.011 & 5.266 $\pm$ 0.024 & 8.779 $\pm$ 0.339 & 0.532 $\pm$ 0.028 & 5.395 & 1.748 $\pm$ 0.106 & 71.63 \\
        4590 & -1.25 $\pm$ 0.08 & 4.821 $\pm$ 0.012 & 5.099 $\pm$ 0.044 & 8.740 $\pm$ 0.077 & 0.472 $\pm$ 0.055 & 4.754 & 1.837 $\pm$ 0.193 & 78.89 \\
        5024 & -1.21 $\pm$ 0.08 & 5.399 $\pm$ 0.010 & 5.588 $\pm$ 0.041 & 11.136 $\pm$ 0.013 & 0.370 $\pm$ 0.061 & 10.567 & 1.361 $\pm$ 0.132 & 213.65 \\
        5053 & -1.26 $\pm$ 0.04 & 4.482 $\pm$ 0.012 & 4.733 $\pm$ 0.115 & 17.621 $\pm$ 0.110 & 0.439 $\pm$ 0.149 & 10.337 & 1.648 $\pm$ 0.439 & 107.33 \\
        5272 & -1.02 $\pm$ 0.08 & 5.369 $\pm$ 0.009 & 5.669 $\pm$ 0.024 & 8.270 $\pm$ 0.016 & 0.498 $\pm$ 0.030 & 6.981 & 1.595 $\pm$ 0.094 & 186.08 \\
        5286 & -0.64 $\pm$ 0.04 & 5.281 $\pm$ 0.010 & 5.579 $\pm$ 0.021 & 5.016 $\pm$ 0.019 & 0.496 $\pm$ 0.027 & 2.789 & 1.129 $\pm$ 0.060 & 143.73 \\
        5466 & -1.14 $\pm$ 0.06 & 4.603 $\pm$ 0.012 & 4.763 $\pm$ 0.090 & 16.192 $\pm$ 0.130 & 0.308 $\pm$ 0.145 & 8.617 & 1.162 $\pm$ 0.243 & 96.28 \\
        5897 & -1.06 $\pm$ 0.14 & 4.844 $\pm$ 0.011 & 5.329 $\pm$ 0.045 & 11.527 $\pm$ 0.157 & 0.672 $\pm$ 0.035 & 8.142 & 2.906 $\pm$ 0.311 & 105.57 \\
        5904 & -0.81 $\pm$ 0.08 & 5.216 $\pm$ 0.010 & 5.615 $\pm$ 0.022 & 6.773 $\pm$ 0.101 & 0.601 $\pm$ 0.022 & 4.603 & 1.863 $\pm$ 0.104 & 106.70 \\
        5986 & -0.58 $\pm$ 0.05 & 5.135 $\pm$ 0.010 & 5.581 $\pm$ 0.042 & 5.377 $\pm$ 0.063 & 0.642 $\pm$ 0.036 & 3.029 & 1.719 $\pm$ 0.171 & 103.72 \\
        6093 & -0.16 $\pm$ 0.05 & 5.053 $\pm$ 0.010 & 5.578 $\pm$ 0.030 & 4.056 $\pm$ 0.067 & 0.701 $\pm$ 0.022 & 1.872 & 1.540 $\pm$ 0.112 & 89.84 \\
        6101 & -1.24 $\pm$ 0.12 & 4.906 $\pm$ 0.095 & 5.222 $\pm$ 0.075 & 13.732 $\pm$ 0.086 & 0.516 $\pm$ 0.135 & 10.199 & 1.773 $\pm$ 0.493 & 84.86               
        \\
        6144 & 0.02 $\pm$ 0.07 & 4.329 $\pm$ 0.010 & 4.344 $\pm$ 0.491 & 5.569 $\pm$ 0.384 & 0.034 $\pm$ 1.092 & 0.939 & 0.416 $\pm$ 0.470 & 21.78 \\
        6218 & -0.36 $\pm$ 0.06 & 4.654 $\pm$ 0.009 & 5.026 $\pm$ 0.031 & 5.232 $\pm$ 0.147 & 0.576 $\pm$ 0.032 & 1.640 & 1.587 $\pm$ 0.118 & 60.57 \\
        6254 & -0.57 $\pm$ 0.10 & 4.937 $\pm$ 0.009 & 5.335 $\pm$ 0.032 & 5.815 $\pm$ 0.093 & 0.599 $\pm$ 0.031 & 2.685 & 1.716 $\pm$ 0.132 & 74.75 \\
        6341 & -0.77 $\pm$ 0.05 & 5.128 $\pm$ 0.010 & 5.513 $\pm$ 0.032 & 5.599 $\pm$ 0.019 & 0.588 $\pm$ 0.032 & 3.269 & 1.571 $\pm$ 0.121 & 149.10 \\
        6362 & -0.58 $\pm$ 0.07 & 4.729 $\pm$ 0.013 & 5.192 $\pm$ 0.035 & 8.456 $\pm$ 0.244 & 0.655 $\pm$ 0.030 & 4.173 & 1.991 $\pm$ 0.171 & 93.25 \\
        6541 & -0.55 $\pm$ 0.05 & 5.064 $\pm$ 0.011 & 5.506 $\pm$ 0.052 & 5.096 $\pm$ 0.071 & 0.639 $\pm$ 0.044 & 2.654 & 1.718 $\pm$ 0.210 & 47.17 \\
        6584 & -0.78 $\pm$ 0.06 & 4.745 $\pm$ 0.010 &                   & 7.760 $\pm$ 0.082 &                   &       &                   & 
        \\
        6723 & -0.16 $\pm$ 0.06 & 4.772 $\pm$ 0.009 & 5.227 $\pm$ 0.034 & 4.989 $\pm$ 0.083 & 0.650 $\pm$ 0.028 & 1.726 & 1.789 $\pm$ 0.144 & 49.12 \\
        6752 & -0.43 $\pm$ 0.08 & 4.996 $\pm$ 0.010 & 5.289 $\pm$ 0.005 & 5.652 $\pm$ 0.118 & 0.490 $\pm$ 0.013 & 2.409 & 1.303 $\pm$ 0.032 & 102.06 \\
        6779 & -0.59 $\pm$ 0.05 & 4.823 $\pm$ 0.010 & 5.384 $\pm$ 0.081 & 6.291 $\pm$ 0.104 & 0.725 $\pm$ 0.052 & 3.305 & 1.955 $\pm$ 0.368 & 119.02 \\
        6809 & -0.83 $\pm$ 0.07 & 4.935 $\pm$ 0.011 & 5.283 $\pm$ 0.028 & 7.289 $\pm$ 0.106 & 0.551 $\pm$ 0.031 & 3.922 & 1.767 $\pm$ 0.122 & 79.93 \\
        7078 & -1.00 $\pm$ 0.04 & 5.448 $\pm$ 0.010 & 5.820 $\pm$ 0.018 & 6.508 $\pm$ 0.015 & 0.575 $\pm$ 0.020 & 5.856 & 1.608 $\pm$ 0.077 & 223.09 \\
        7089 & -0.72 $\pm$ 0.06 & 5.415 $\pm$ 0.008 & 5.954 $\pm$ 0.029 & 6.127 $\pm$ 0.015 & 0.711 $\pm$ 0.020 & 5.495 & 2.238 $\pm$ 0.155 & 287.75 \\
        7099 & -0.80 $\pm$ 0.03 & 4.771 $\pm$ 0.013 & 5.136 $\pm$ 0.026 & 5.369 $\pm$ 0.088 & 0.569 $\pm$ 0.029 & 2.198 & 1.620 $\pm$ 0.109 & 136.80 \\
        \hline  
    \end{tabular} 
\end{table*}

\cite{sollima2012}, \cite{sollima2017a} and \citeauthor{sollima2017b} have described
a method to determine the global GCs MF and the 
structural/dynamical 
parameters of GCs and we adopt their method in this paper.

Briefly, an iterative procedure has been implemented. At each iteration, 
a guess of the MF and of the model parameters is chosen and a synthetic population of particles 
with the corresponding masses (magnitudes) and radial distributions has been 
simulated.
The effect of completeness and photometric errors are included using the artificial 
star experiments described in Sect. \ref{Sec:2} and a mock catalog of masses and distances is created.
The corrections to the guess MF are calculated by comparing the distribution of 
synthetic and observed stars in the mass-distance plane, and the updated MF is 
used as input for the next iteration. This algorithm converges after a few iterations providing the 
output global MF.

Specifically, the algorithm can be schematically summarized as follows:

\begin{enumerate}

\item{An initial guess of the MF (defined by the coefficients $k_{i}$) is 
made. In the present analysis the coefficients corresponding to a \cite{kroupa2001} 
IMF have been adopted as first guess.}

\item{A synthetic stellar population has been created
by extracting $10^{6}$ stars from the adopted MF with masses between 0.1 and
8 M$_{\odot}$. It has been assumed that the mass loss of stars with 
masses $<M_{\text{tip}}$ (the mass at the tip of the red giant branch, RGB)
is negligible while those with masses $>M_{\text{tip}}$ have been evolved into
white dwarfs (WDs) following the initial-final mass relation of \citet{kaliari2009}
\begin{equation}
m_{\text{WD}}=0.109\,m+0.428.
\end{equation}     
According to the upper limit of the IMF, the white dwarfs are all retained by the GC 
unlike the other types of compact remnants. This implicitly assumes that the neutron stars
and the black holes are ejected during the GC evolution by natal kicks and/or
the Spitzer instability \citep{spitzer1987}.
The fraction of mass in remnants in each bin ($\nu_{i}$) has also been calculated.
No binaries have been simulated. These objects constitute only a small ($<5\%$) 
fraction of objects in GCs \citep{milone2012} and their presence is not expected to 
significantly affect the final MF.}

\item{The corresponding V and I magnitudes of the synthetic 
stars have been derived by interpolating the masses of visible stars
through the mass-luminosity relation of the adopted isochrone from 
the \cite{dotter2007} database, adopting the best fit metallicity, age, distance modulus 
and reddening provided by \cite{dotter2010} (see section \ref{Sec:2}).
Zero luminosity in both bands has been assumed for the population of remnants.
Note that we used the same isochrone, distance modulus 
and reddening for each GC, both in photometric data reduction and in MF analysis processes.}

\item{A synthetic population of field stars has been derived from
the Galactic model provided by \cite{robin2003} covering an
area of 1 sq. deg around each GC centre and the V and I 
magnitudes have been transformed into the HST/ACS photometric system 
using the transformation provided by 
\citet{sirianni2005}.}

\item{We defined eight evenly-spaced mass groups ranging
from $0.1 \, \text{M}_\odot$ to $M_{tip}$ and one additional bin for stars more
massive than $M_{tip}$ including massive WDs. Eight I-band
magnitude intervals containing the above defined mass bins 
(excluding the one related to remnants) have been also 
defined accordingly.
Real, field and synthetic stars with colours within three times the mean 
locus of MS stars, have been binned in these groups.
The projected density of field stars in each bin $\Gamma_{i}^{\text{field}}$ is calculated by dividing the 
number of field stars contained in each mass bin by the area of the extracted 
field catalog (1 sq. deg).} 

\item{The completeness factor, $C(m,R)$, defined as the fraction
of artificial stars recovered in each magnitude (mass) bin and contained in concentric 
annular regions of 0.1 arcmin width, has been estimated.}

\item{The azimuthal coverage of the observational
field of view as a function of the projected distance, $Az(R)$, has been calculated.
}

\item{We compute the value of the 
log-likelihood 
function defined as:
\begin{equation}
log~L=\sum\limits^{N^{\text{obs}}_{\text{tot}}}_{j=1} \log [P(m_j,R_j)],
\end{equation}
where $N^{\text{obs}}_{\text{tot}}$ is the total number of observed stars, $m_{j}$ and $R_{j}$ are the 
mass and projected distance of the $j-$th observed star and 
$P(m_j,R_j)$ is the total probability density function to find a star
with mass $m_j$ at a projected distance from the centre $R_j$.
This last function is calculated as the model density 
in the $m-R$ plane, after correcting for completeness and azimuthal coverage
\begin{multline}
P(m_j,R_j)=\left[\frac{N_{\text{tot}}^{\text{obs}}-N_{\text{tot}}^{\text{field}}}{\sum\limits^{H}_{i=1}N_{i}}\Gamma(m_j,R_j)(1-\nu_{j})+\Gamma_{j}^{\text{field}}\right]~ \\
\times\frac{C(m_j,R_j)~Az(R_j)}{N_{tot}^{obs}},
\end{multline}
where $N_{i}$ and $\Gamma(m_j,R_j)$ are calculated from eq.s \ref{num-eq} and 
\ref{surf-den-eq}, $\Gamma_{i}^{\text{field}}$ is the projected 
number density of field star in the same mass bin of the $j$th star, $N_{tot}^{obs}$ is the total number of stars in the observed catalog and
\begin{equation*}
N_{\text{tot}}^{\text{field}}=\sum\limits^{H}_{i=1}\Gamma_{i}^{\text{field}}\int_{0}^{r_{t}}2\pi R~C(m_i,R)~Az(R) dR.
\end{equation*}
}
\begin{figure}
\begin{center}
\includegraphics[width=85mm]{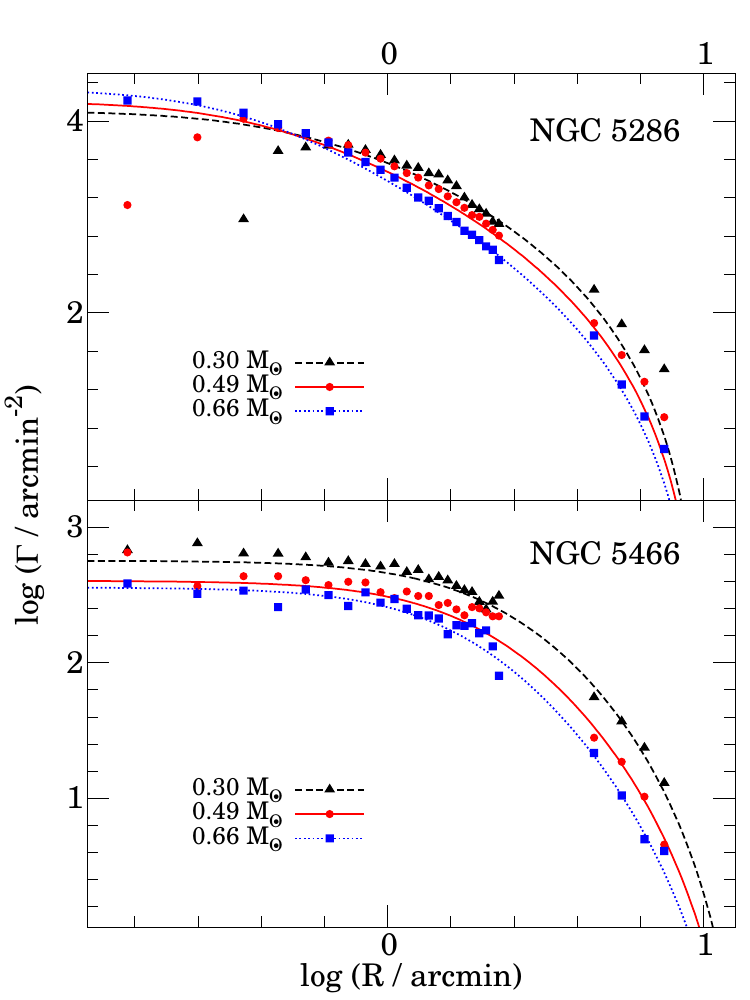}
\caption{Projected density profiles for three mass groups in two GCs:
NGC 5286 (top panel) and NGC 5466 (bottom panel). Coloured dots mark completeness-corrected densities of different mass groups. 
The density profile of the best-fitting models 
are shown with solid, dotted and dashed lines. The same plot for the entire sample of 
GCs is available as supplementary material in the online version of the paper.}
\label{dR}
\end{center}
\end{figure} 

\begin{figure*}
\begin{center}
\includegraphics[width=120mm]{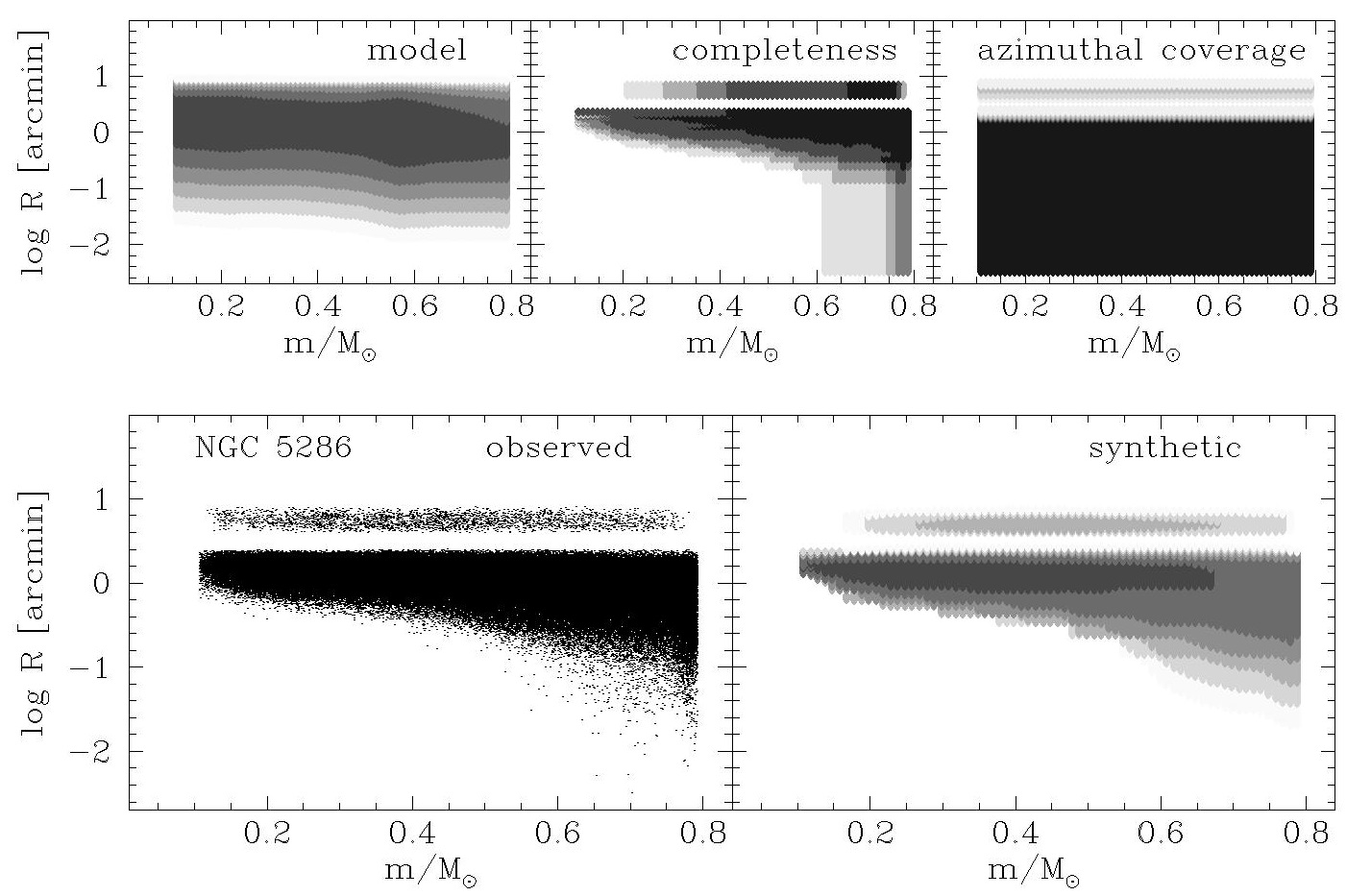}
\caption{Distribution of stars in the
$m-\log R$ plane for NGC 5286 (bottom-left panel).
The probability density predicted by 
the corresponding best-fitting final models (bottom-right panel),
initial model prediction (top-left panel),
the completeness (top-middle panel),
and the azimuthal coverage (top-right panel)
are shown.}
\label{mR}
\end{center}
\end{figure*} 
\item{For the guess choice of $k_i$, the $W_{0},~r_{c}$ space has been searched 
to find the values maximizing the above merit function using a Powell's direction set algorithm 
\citep{brent1973}}. 

\item{For the best fit pair of ($W_{0},~r_{c}$) values, the corresponding multimass model has 
been computed. Synthetic stars have been distributed according to the density 
profile of their correspondig mass bin across the field of view.}

\item{For each synthetic star, a particle from the artificial star library
with I magnitude within 0.25 mag and the distance from the GC centre 
within 0.1 arcmin,
with respect to the same quantities of the given star
 has been selected and, if recovered (see section \ref{Sec:2}),
its output-input magnitude and colour shift 
have been added to those of the corresponding star.
The same procedure has been applied to field stars.} 

\item{For each synthetic star located at a distance $R_j$ from the centre, 
a random number $\eta$ uniformly distributed between 0 and 1 
has been extracted and the star is rejected if $\eta>Az(R_j)$.
The same procedure has been applied to field stars.  
At the end of this step a mock catalog of synthetic stars 
corresponding to the guess choice of $k_{i},~W_{0}$ and $r_{c}$ is obtained, 
accounting for the photometric errors, incompleteness and azimuthal coverage.
}

\item{The number of stars in the observed ($N^{obs}_{i}$), field ($N^{field}_{i}$) 
and synthetic ($N^{mock}_{i}$) 
catalogs contained in the eight I magnitude bins have been counted and the 
$k_{i}$ coefficients have been updated using the relation
\begin{equation}
k_{i}'=k_{i}\frac{N_{tot}^{mock}~(N^{obs}_{i}-N^{field}_{i})}{(N_{tot}^{obs}-N_{tot}^{field})~N^{mock}_{i}}
\end{equation}
}

\end{enumerate}

The entire procedure has been repeated iteratively until convengerce.
The global MF is determined by the relative number of stars in the eight mass bins
in the last iteration (eq. \ref{num-eq}).

The best fit of the projected density profiles 
for two GCs
and of the distribution of stars 
in the $m-R$ plane 
for NGC 5286
are shown in Figs. 
\ref{dR} and \ref{mR}, as an example. According to Fig. \ref{mR}, the density contours
of the best-fitting model are generally in good agreement with the distribution of stars in the $m-R$ plane.
As an alternative view, in Fig. \ref{dR}, the surface density predicted by the best-fitting model
for three mass groups is overplotted to the observational density profile for the same GCs. 
Note that, while the model captures the general behaviour of mass-segregation, some 
discrepancies are apparent inside the core of these GCs.

The stellar mass of each GC has been derived by normalizing
the total number of stars in the best-fitting model
to those counted in the observed catalogue 
\begin{equation}
M_{\text{lum}}=\frac{N_{\text{tot}}^{\text{obs}}-N_{tot}^{field}}{N_{\text{tot}}^{\text{mock}}} \sum\limits^{H}_{i=1} N_i m_i (1-\nu_i), 
\end{equation}

The dynamical mass has been estimated by comparing the observed velocity dispersion profile with
the prediction of the best-fit model.
In particular, the mass of the model has been chosen as the one minimizing the penalty function:
\begin{equation}
\mathcal{L}= \sum\limits^{N}_{j=1} \Bigg \lbrace \frac{(v_j-\bar{v})^{2}}{\sigma_{\text{LOS},8}^{2} (R_j)+\epsilon_{j}^{2}}+\ln \big[\sigma_{\text{LOS},8}^{2} (R_j)+\epsilon_{j}^{2} \big] \Bigg \rbrace,
\end{equation}
where $N$ is the number of radial velocities, $v_j$ is the radial velocity of the $j$th star, $\bar{v}$ is 
the mean velocity of the sample, $\epsilon_{j}$ is the uncertainty
of the radial velocity and $\sigma_{\text{LOS},8}(R_j)$ is the 
line-of-sight velocity dispersion predicted by
the best-fitting model at the projected distance $R_j$ of the $j$th star
for the eighth mass group (see eq. \ref{los-vel-eq}). 
The choice of the 8-th bin is justified by the fact that 
the considered radial velocity catalogues contain
stars along the RGB covering a limited range of masses. 

The total V-band luminosity has been calculated using 
equation \ref{lv-equ}, allowing to derive the
mass-to-light ratio, $M_{\text{dyn}}/L_V$. 

The fraction of dark remnants has been estimated using the following relation:
\begin{equation}
f_{\text{remn}}=1-\frac{{M_\text{lum}}}{{M_\text{dyn}}}
\end{equation}

The half-mass radius, $r_h$, has been evaluated as the radius including
half of the GC mass while the half-mass relaxation time as \citep{spitzer1987}
\begin{equation}
t_{rh}=0.138\dfrac{{M_\text{dyn}^{1/2}}r_h^{3/2}}{G^{1/2}\bar{m} \ln(\gamma M_\text{dyn} / \bar{m})}
\end{equation}   
where $\gamma=0.11$ \citep{giersz1996} and $\bar{m}$ is the mean mass of stars (eq. \ref{m-eq}).

\section{Results}\label{Sec:5}

\subsection{An Overview of GCs Mass Function and Their Parameters}

\begin{figure*}
\begin{center}
\includegraphics[width=180mm]{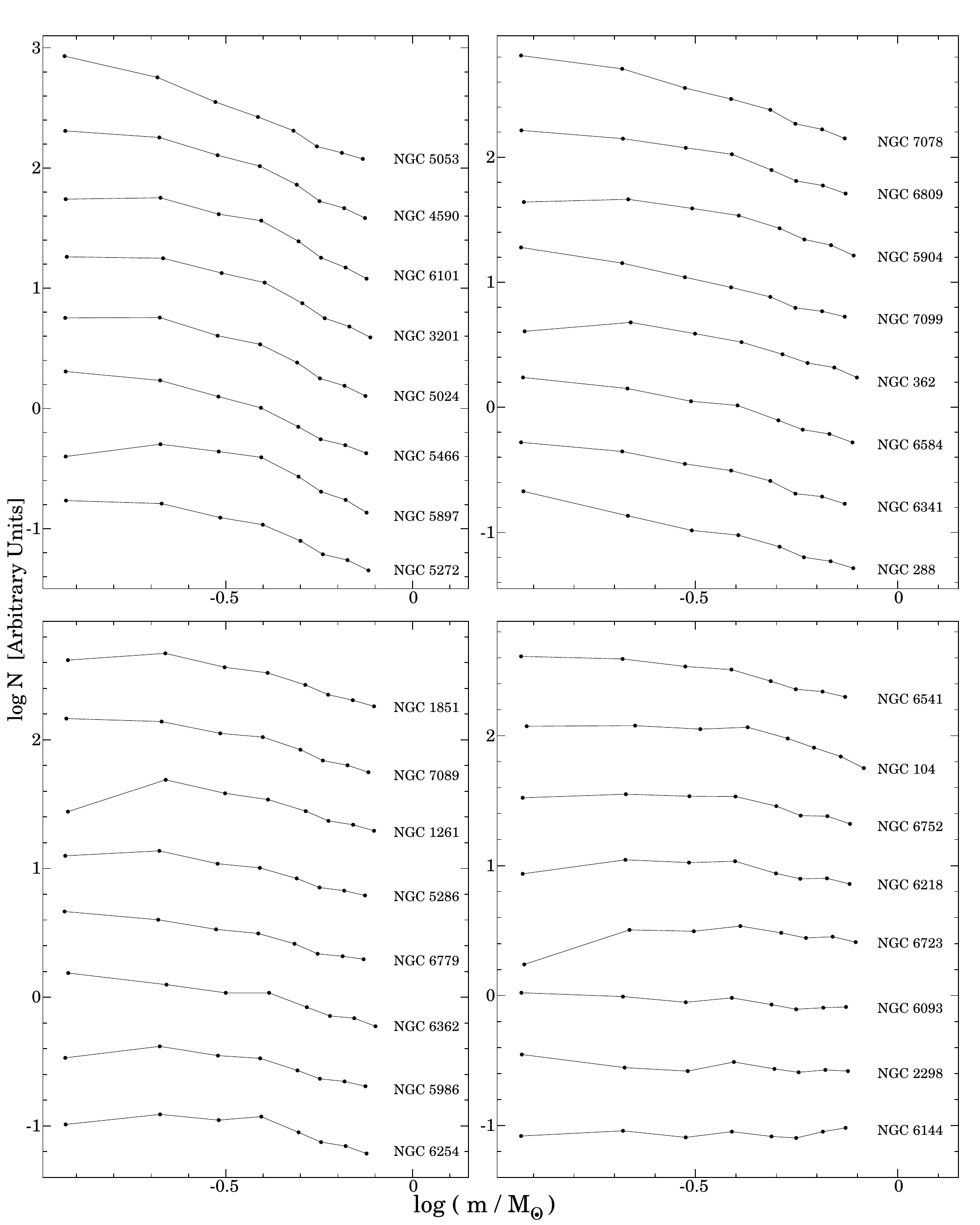}
\caption{Present-day mass function of the 32 GCs in our sample. A vertical
shift has been added to each cluster in each panel for clarity. From left to right panel
and top to the bottom panel, GCs have been sorted according to their MF slope, in descending order.}
\label{mf}
\end{center}
\end{figure*}

The derived global MFs and the 
structural/dynamical 
parameters of 32 GCs are listed
in Table \ref{table_main} and the present-day MFs of these GCs are shown in Fig 
\ref{mf}. The MF slopes $\alpha$ calculated by fitting a single
power-law, $F(m)\propto m^{\alpha}$, in the mass range $0.2<m/M_{\odot}<0.8$ 
have also been calculated. 

In the considered sample, the derived parameters cover wide ranges in MF slope 
($-1.26<\alpha<0.02$), mass ($2.2\times10^{4}<M/{\text{M}}_{\odot}<10^{6}$), 
mass-to-light ratio ($0.4<M/L_{V}<2.9$) and half-mass radius ($4<r_{h}/\text{pc}<17.6$).
For reference, a \cite{kroupa2001} IMF in this mass range has an average slope of -1.567.
As already found in \citeauthor{sollima2017b}, GCs with a steeper MF slope tend to have a 
convex shape with a systematic depletion of stars at $M<0.2~M_{\odot}$, while the MF of 
GCs with a flatter MF are much better fitted by single power-laws. 

\subsection{Dependence on assumptions: NGC 104 as test case}

\begin{figure}
\begin{center}
\includegraphics[width=85mm]{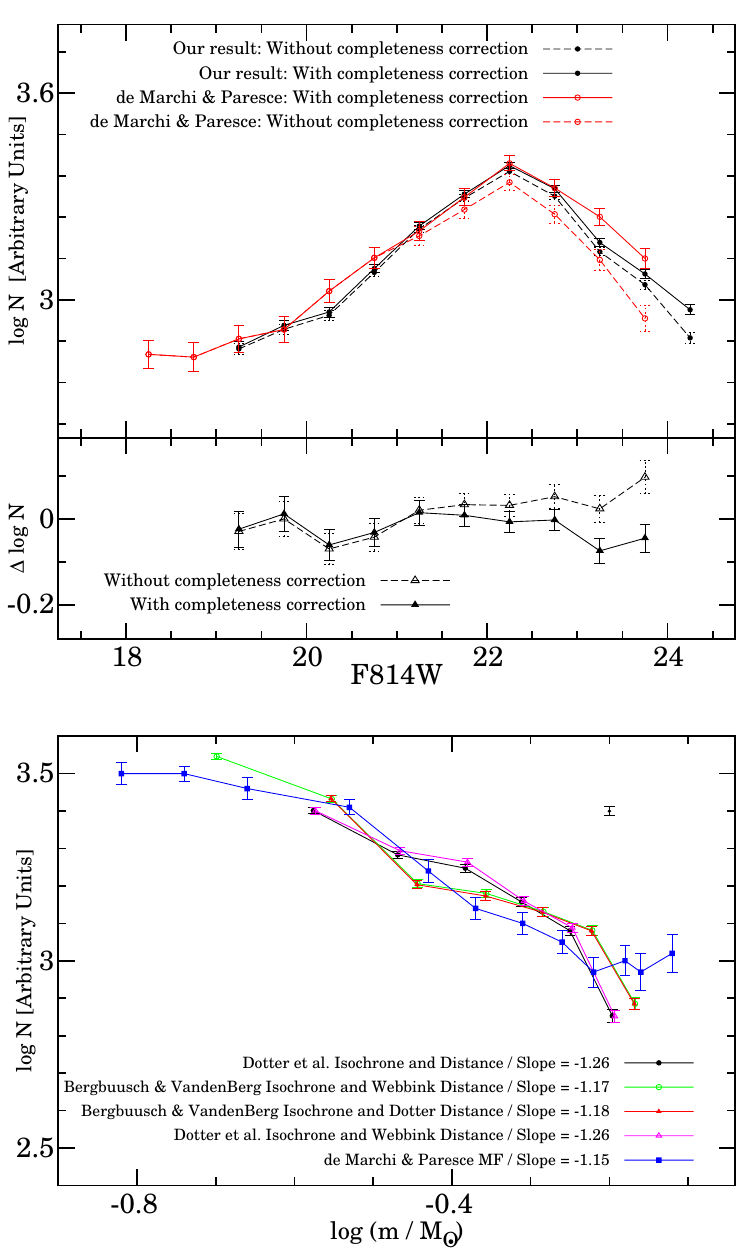}
\caption{Comparisons between the luminosity function (top panel) and MF (bottom panel) of our 
analyzed P2 region against \protect\cite{demarchi1995} region in NGC 104. The difference of the
luminosity function between two works calculated for raw and completeness corrected 
luminosity functions is shown in the lower box of the top panel. 
The MF slope reported in the bottom panel have been calculated in the mass range $-0.60<\log (m/M_{\odot})<-0.15$. 
The error bars are related to Poisson errors estimated from counting statistics. 
An arbitrary shift has been added to
each MF and luminosity function for clarity.}
\label{MF_LF}
\end{center}
\end{figure}

The MF of the GC NGC 104 has been studied by \cite{demarchi1995} who used 
HST/WFPC2 observations at 4.6 arcmin from the centre through F606W and F814W filters i.e. 
similar to one of the ACS pointing used in the present analysis (P2 in Fig. \ref{map}; at 
$\sim 6\arcmin$), adopting a different distance and mass-luminosity relation.
The MF slope calculated from this work in the range $-0.60<\log (m/M_{\odot})<-0.15$ turns out to be 
$\alpha=-1.15$, flatter than what is estimated in the present work ($\alpha=-1.26$).  
This gives the opportunity to test the effect of the various assumptions on the derived MF 
providing also a validation of the photometric analysis and completeness correction with a 
completely independent study.
Note that our analyzed region
is located on the opposite side of the \cite{demarchi1995} field with respect to the 
cluster centre and is larger. In the top panel of Fig. 
\ref{MF_LF} the luminosity functions of \cite{demarchi1995} and our "P2" region 
are compared, with and without applying the completeness correction.
A good agreement between the two works is apparent, with only random 
fluctuations of $<0.1~\text{dex}$ amplitude and no systematic trend. This is 
particularly apparent when completeness correction are applied in both works, accounting 
for the different depth of the two different observations. This agreement 
indicates that the difference in the MF slope between these two works does not 
depend on either the photometry or on the completeness, but on the different 
assumptions on distance and/or mass-luminosity relation.

\cite{demarchi1995}
derived the MF of stars within their region using a \cite{bergbusch1992} 
isochrone with [Fe/H]=-0.65 and age of 14 Gyr for NGC 104 and adopt a distance 
of 4.6 kpc \citep{webbink1985}, while in the present work we use a 
\citet{dotter2007} isochrone with [Fe/H]=-0.7 and age of 12.75 Gyr and adopt a 
distance of 4.5 kpc. 

To check the effect of such assumptions we derived the MF of region P2 using 
all the combinations of isochrones \citep{dotter2007,bergbusch1992}
and distance moduli \citep{dotter2007,webbink1985},
and compared these four resulting MFs with the MF derived by 
\cite{demarchi1995} in their analysed field. 
In the bottom panel of Fig. \ref{MF_LF} this comparison is shown.
It is clear that the small difference in the adopted distance produces a 
negligible effect across the entire mass range. 
A larger effect is produced by the different mass-luminosity relation, 
with the \cite{bergbusch1992} isochrone providing a flatter MF than that derived 
using \cite{dotter2007} models ($\Delta \alpha\sim 0.1$).
So, this effect can entirely explain the difference in the measured 
slope.

Summarizing, at least for the case of NGC 104, the impact of assumptions on 
distance and adopted models should not produce an effect exceeding 
$\Delta\alpha\sim 0.15$ in the MF slope.

\subsection{Comparison with previous works}

\begin{table*}
    \centering
    \caption{The mean $(\mu)$ and standard deviation $(\sigma)$ of differences between this work and the others
    in estimated MF slopes (third column), dynamical masses (fourth column), half-mass radii (fifth column), and 
    mass-to-light ratios (sixth column).  }
    \label{table3}
    \begin{tabular}{cccccccccc}
        \hline 
        \hline    
        & Statistic & $\Delta \alpha$ & $\Delta \log (M/\text{M}_{\odot})$ & $\Delta \log (r_h/\text{pc})$ & $\Delta [(M/\text{M}_{\odot})/(L/\text{L}_{\odot})]$  \\
        \hline
        \hline  
        This Work - \cite{paust2010} & $\mu$ & $0.23\pm 0.12$ & & &  \\
        & $\sigma$ & 0.43 & &  \\
        \hline
         This Work - \citeauthor{baumgardt2018} & $\mu$ &  & $0.04\pm 0.02$ & $0.10\pm 0.01$ & $-0.27\pm 0.08$ \\
        & $\sigma$ &  &  0.11 & 0.08 & 0.46 \\
        \hline
       This Work - \citeauthor{sollima2017b}  & $\mu$ & $0.03\pm 0.03$ & $0.005\pm 0.020$ & $0.02\pm 0.02$ & $-0.10\pm 0.06$ \\
        & $\sigma$ & 0.16 &  0.10 & 0.10 & 0.31 \\
        \hline 
        \hline               
    \end{tabular}   
\end{table*}     

\begin{figure*}
\begin{center}
\includegraphics[width=105mm]{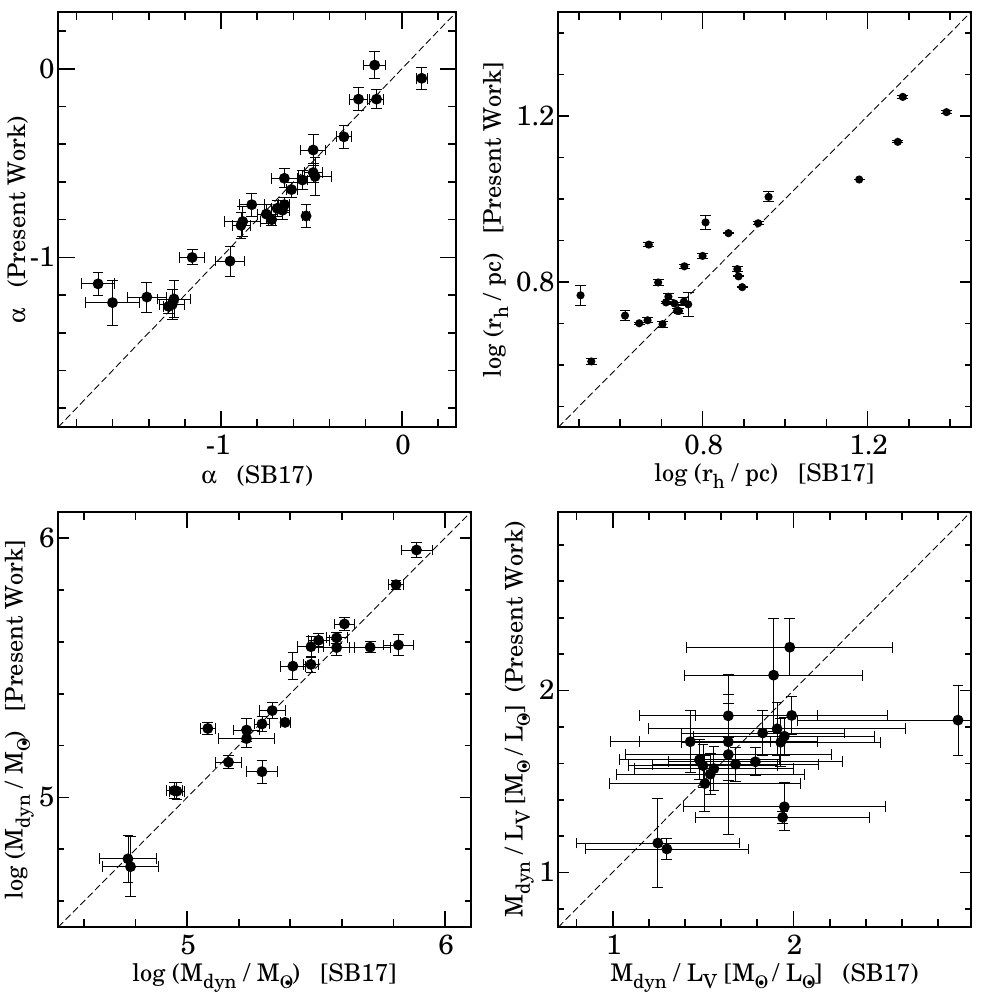}
\caption{Comparison between the MF slopes (top-left panel), half-mass radii (top-right panel),
dynamical masses (bottom-left panel) and mass-to-light ratios (bottom-right) derived in the present work and
those determined by \protect\citeauthor{sollima2017b}. The
one-to-one relation is marked by the dashed line.} 
\label{comparison}
\end{center}
\end{figure*}
\begin{figure*}
\begin{center}
\includegraphics[width=180mm]{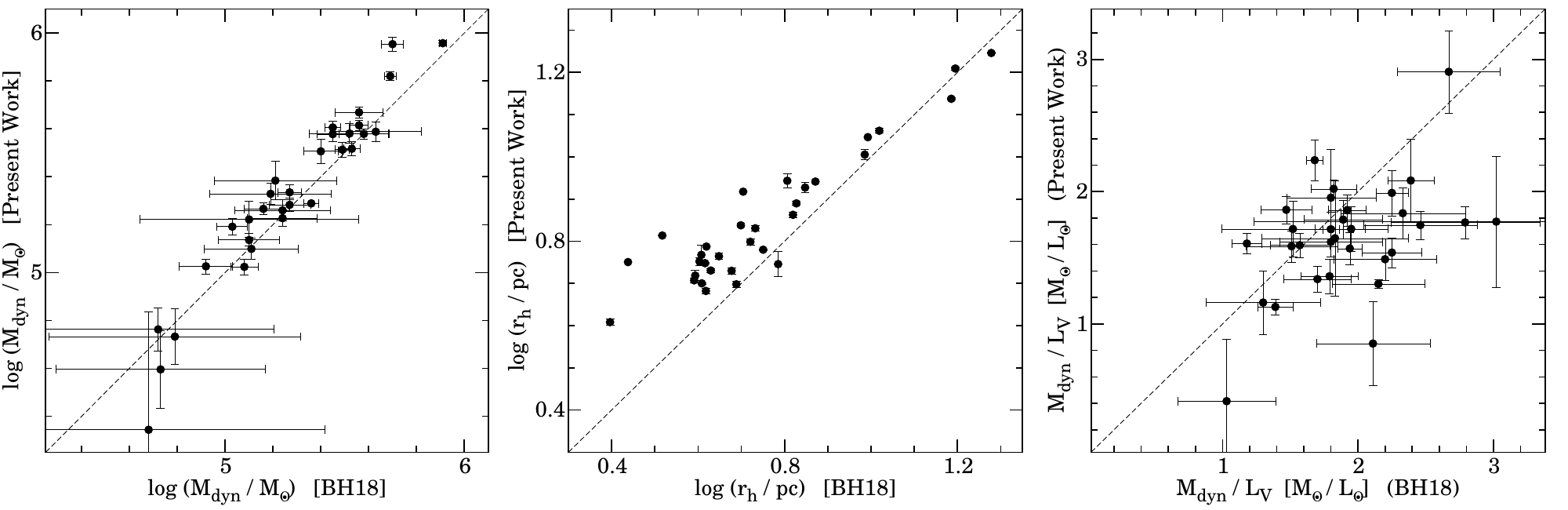}
\caption{Comparison between the dynamical masses (left panel), half-mass radii (middle panel)
and mass-to-light ratios (right panel) derived in the present work and
those determined by \protect\citeauthor{baumgardt2018}. The
one-to-one relation is marked by the dashed line.} 
\label{comparison2}
\end{center}
\end{figure*}

In this section we compare the results obtained in this analysis with 
those obtained by \citeauthor{sollima2017b} and \cite{paust2010} (who applied a similar technique using 
only data in the central cluster region) and with those by \citeauthor{baumgardt2018}.
The statisics of the differences between our work with those mentioned 
above have been summarized in Table \ref{table3}.

In Fig \ref{comparison}, the MF slopes and half-mass radii of 28 GCs
and the dynamical masses and the mass-to-light ratios of 23 GCs in common between this work and 
\citeauthor{sollima2017b}
are compared. 
The largest differenes in the estimated MF slopes ($\Delta \alpha>0.2$)
are related to the two GCs which have the steepest MF among 28 GCs in 
\citeauthor{sollima2017b} 
(e.g. NGC 5466 and NGC 6101) while they have flatter MF in the present work. 
The mean differences between the two studies 
indicate good agreements in estimated MF slopes, dynamical masses and mass-to-light ratios.
The difference between two studies in estimated half-mass radii shows 
that the three GCs with the largest difference between the 
half-mass radii estimated in these works i.e. NGC 6101, NGC 5466, NGC 5024, are 
those with the largest difference in the estimated MF slope. 
This indicates the importance of the MF constraint outside the ACS field of 
view for these extended clusters.
  
Fifteen GCs are in common with the work by \cite{paust2010}. 
While the difference between these two works is 
not statistically significant, our MFs are on average flatter than those measured by \cite{paust2010}.
These authors used the same HST/ACS photometric data in the central 
region and the same multimass model fitted in an annular region close to the half-light radius.

In Fig \ref{comparison2}, the half-mass radii of 32 GCs,
masses and mass-to-light ratios of the 31 GCs in common between the present work and 
\citeauthor{baumgardt2018} are compared.
These authors determined the masses and structural/dynamical parameters of GCs 
by fitting a large set of N-body simulations to the velocity dispersion profile from the same set of radial velocities adopted in this work. 
The mean differences in
 the estimated masses and $M/L$ ratios 
indicate
good agreement within
the errors in spite of the difference between our adopted multimass models and N-body simulation.
The estimated half-mass radii are instead slightly larger in this work.
This difference arises mainly from the more compact GCs, where the constraint provided by 
the external field photometry tends to favour a slightly larger mass at large radii.

\subsection{Mass segregation}
\begin{figure}
\begin{center}
\includegraphics[width=85mm]{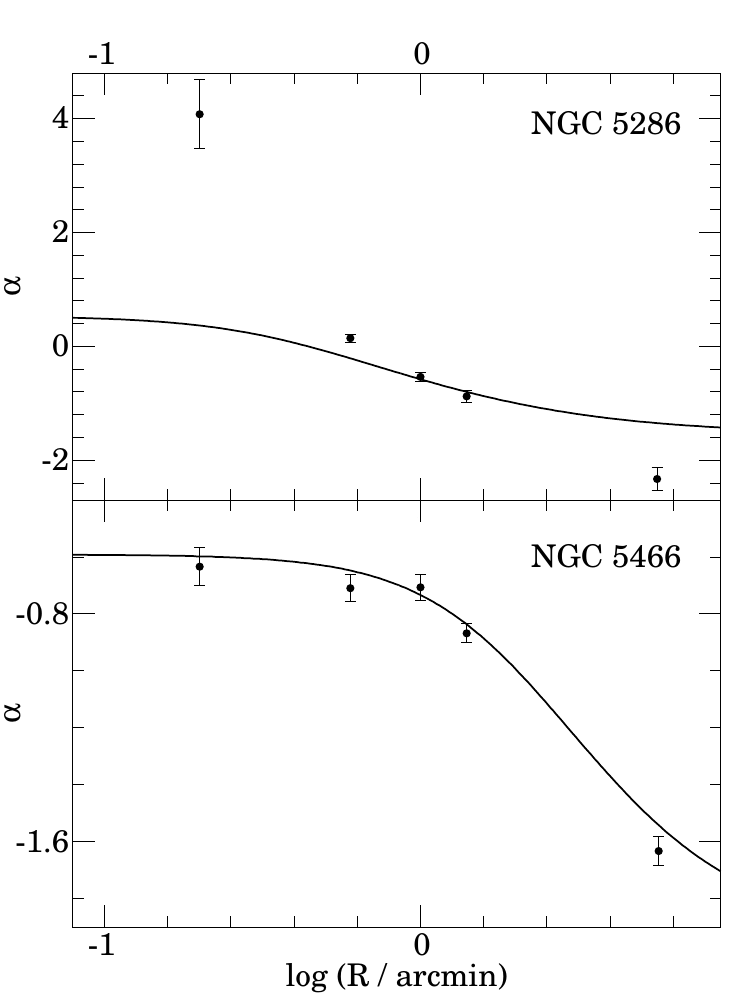}
\caption{Radial variation of the MF slopes of two GCs: NGC 5286
(top panel) and NGC 5466 (bottom panel). The slope predicted by the corresponding
best-fit model are shown by solid lines.}
\label{alphaR}
\end{center}
\end{figure} 

\begin{figure*}
\begin{center}
\includegraphics[width=120mm]{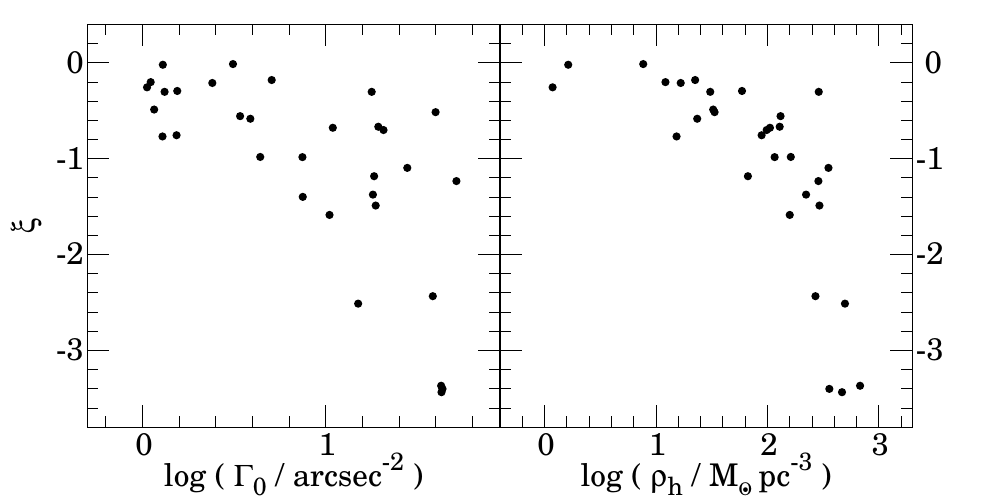}
\caption{Projected central density (left panel) and half-mass density 
(right panel) vs. $\xi$ for all GCs in our 
sample.}
\label{dalpha_rho}
\end{center}
\end{figure*} 

The MFs derived here rely on the assumption that King-Michie multimass models 
\citep[computed according to the prescriptions of ][]{gunn1979} provide an 
adequate description of the mass segregation occurring in GCs.
It is interesting to check this hypothesis by comparing the predicted and observed 
variation of the MF slope at different distances from the cluster center.

For this purpose, we considered for each GC five regions at different distances from the cluster center: 
four annular regions of $0.4\arcmin$ width in the central field and a single bin containing the parallel pointings. 
Then We computed the completeness-corrected
MF slopes at these radial bins.
The radial behaviour of MF slopes measured in two GCs (NGC 5286 and NGC 5466) 
are compared with those predicted by the best-fitting model in Fig. \ref{alphaR}. 
It is clear that the model
reproduces well the observed radial trend of the MF slope for NGC 5466, while it does not
match the MF slope variation in NGC 5286. 
In particular, in this last GC the observed MF slope varies much more quickly (by 
$\Delta \alpha\sim -6$) 
with radius than what is predicted by the best fit model ($\Delta \alpha\sim -1.2$).
In other words, NGC 5286 appears to be over-segregated with respect to the prediction of the 
King-Michie model.

To quantify the discrepancy between models and observations, for each GC we performed a 
least-squares fit to the residuals of the best fit MF slope ($\Delta \alpha$)
as a function of the distance from the center (in units of $r_{h}$) and adopted 
the slope of this linear fit $\xi\equiv \text{d}\Delta\alpha / \text{d}log~R$ as an indicator of the 
observed over-segregation. 
The over-segregation parameter measured in the whole GCs sample is plotted as a function of the 
projected central density and half-mass density in Fig. \ref{dalpha_rho}.
Both densities clearly correlate with $\xi$ such that, while at relatively small 
densities models correctly reproduce the observed degree of mass segregation ($\xi\sim 0$), 
denser GCs show significantly steeper gradients of $\alpha$ ($\xi<0$).
Note that $\xi$ does not seem to significantly correlate with other quantities. 
This evidence can be interpreted in two alternative ways: {\it i)} the estimated 
completeness is spuriously high in the extremely crowded conditions occurring 
in dense GCs, or {\it ii)} This is a real effect such that
the densest GCs with shorter central relaxation time reach a degree of mass 
segregation higher than what is predicted by multimass models.
In this context, it is interesting to note that the correlation with the central 
projected density (an indicator of the maximum crowding) has a larger spread 
than that with the half-mass density (which is an intrinsic parameter of the GC).

Note that the global MF slopes (listed in Table \ref{table_main}) are strongly constrained by the 
radial interval close to the 
half-mass radius containing the largest fraction of stars. 
So, while any inadequacy of the adopted models reflects in 
the accuracy of the estimated MF slopes, no significant 
systematic errors are expected. This fact is supported by the simulations of
\cite{baumgardt2003} 
who found that the global and local MFs at around 60 per cent of the half-light
radius are approximately the same.

\subsection{Tidal Radii of GCs} 

\begin{figure}
\begin{center}
\includegraphics[width=85mm]{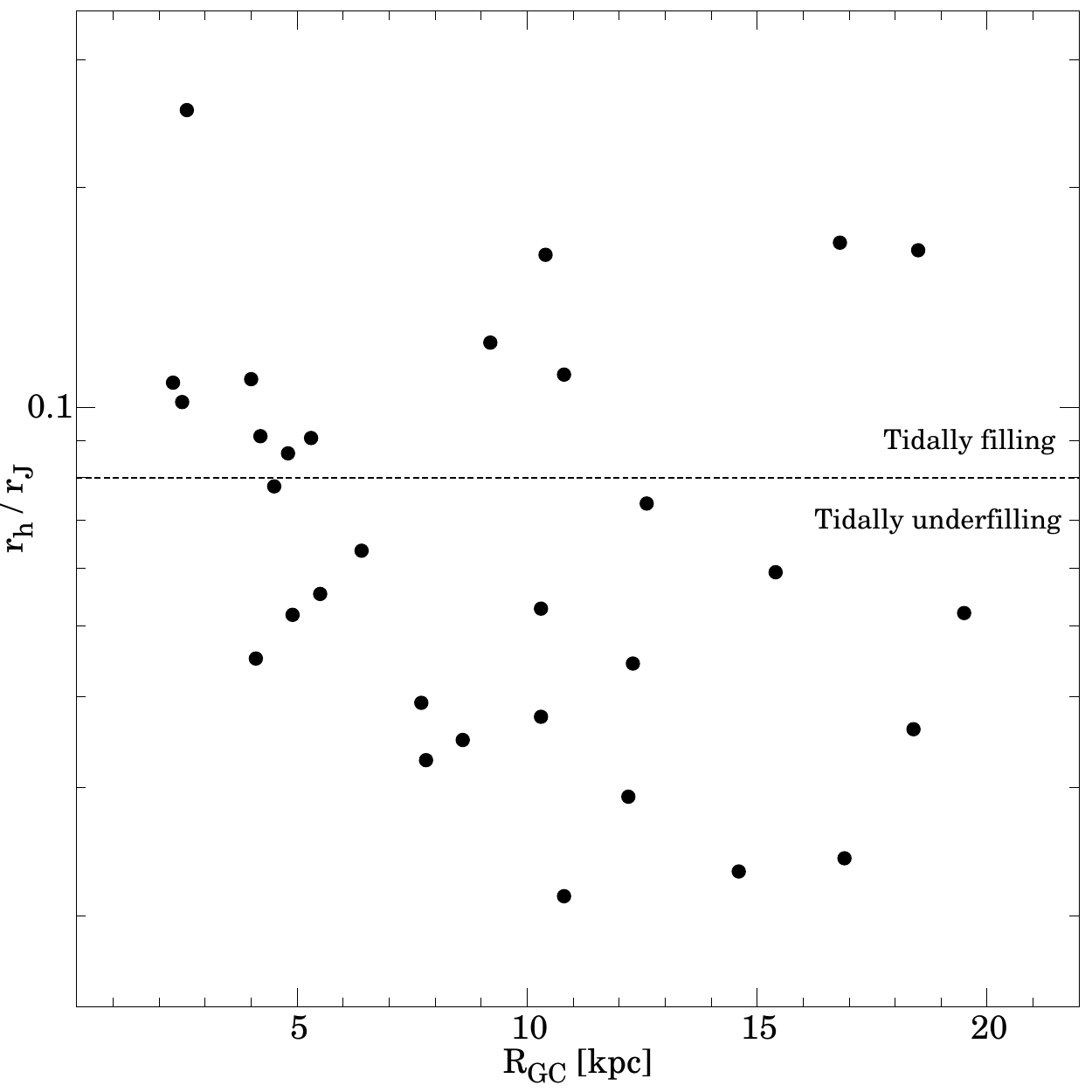}
\caption{Ratio of half-mass radius to Jacobi radius ($r_h/r_J$)
versus Galactocentric distance for 31 Galactic GCs. The dashed line 
empirically separates tidally underfilling
from tidally filling GCs.}
\label{filling}
\end{center}
\end{figure} 
 
We calculated the present-day Jacobi radius ($r_J$; defined as the distance of the inner 
Lagrangian point from the cluster center) of each GC in our sample using the same
formulas described in \cite{sollima2014} (see their appendix A2). 
For this purpose, we used the dynamical masses estimated here and the 
velocities from \emph{Gaia} DR2 determined by \cite{baumgardt2019}. 
We have assumed
a three-component Galactic potential given by the superposition of (i) a Hernquist bulge
with $c=0.7\, \text{kpc}$ and $M_b=1.3\times 10^{10} {\text{M}}_{\odot}$,
(ii) a Miyamoto \& Nagai disk with $a=6.5\, \text{kpc}$,
$b=0.26\, \text{kpc}$ and $M_d=1.085\times 10^{11} {\text{M}}_{\odot}$,
(iii) a logarithmic halo with $v_0=165\, \text{km/s}$ and
$d=12\, \text{kpc}$ (in analogy with the potential adopted by \cite{johnston1995},
with a slightly different normalization of the components to best-fit 
the post-Gaia Galactic kinematics of GCs; Sollima et al., in prep.).

In Fig \ref{filling}, the ratio of the half-mass radius to the Jacobi radius 
as a function of Galactocentric distance derived from \citet[][2010 edition]{harris1996} is shown. 
Among the innermost GCs with $R_{\text{GC}}<8\, \text{kpc}$, two extreme cases 
are notable: (i) NGC 6144
has the highest ratio $(r_h/r_J=0.26)$ because it is the lightest
and one of the nearest GCs from the centre of the Galaxy, (ii) The lowest
ratio belongs to NGC 104 with $r_h/r_J=0.03$ which is the
most massive GC in our sample and the most distant GC from the 
centre of the Galaxy within the mentioned interval.
Generally, there is no clear relationship between $r_h/r_J$ and $R_{\text{GC}}$
in this interval. For outer GCs with $R_{\text{GC}}>8\, \text{kpc}$,
there are two recognizable groups: (i) 5 GCs with $r_h/r_J>0.08$, 
i.e. NGC 3201, NGC 4590, NGC 5053, NGC 5466 and NGC 6101 whose mean mass and
mean half-mass radii are $M_{\text{dyn}}\approx 10^{5} \,{\text{M}}_{\odot}$ and 
$r_{h}\approx 13 \,\text{pc}$ respectively, so they are the extended
and tidally filling GCs. (ii) 12 GCs with $r_h/r_J\lesssim 0.08$. 
Their mean mass and mean half-mass
radii are $M_{\text{dyn}}\approx 3.2 \times 10^{5} \,{\text{M}}_{\odot}$ and
$r_{h}\approx 7 \,\text{pc}$, respectively. Most of the GCs in this group
are compact and tidally underfilling. 

The GCs in the former
group experience a stronger tidal field and are expected to dissolve faster than 
the ones in the latter group. The existence of these two distinct
groups confirms the previous findings by \cite{baumgardt2010}. 
        
\subsection{Analysis of Correlations}
 
\begin{figure*}
\begin{center}
\includegraphics[width=150mm]{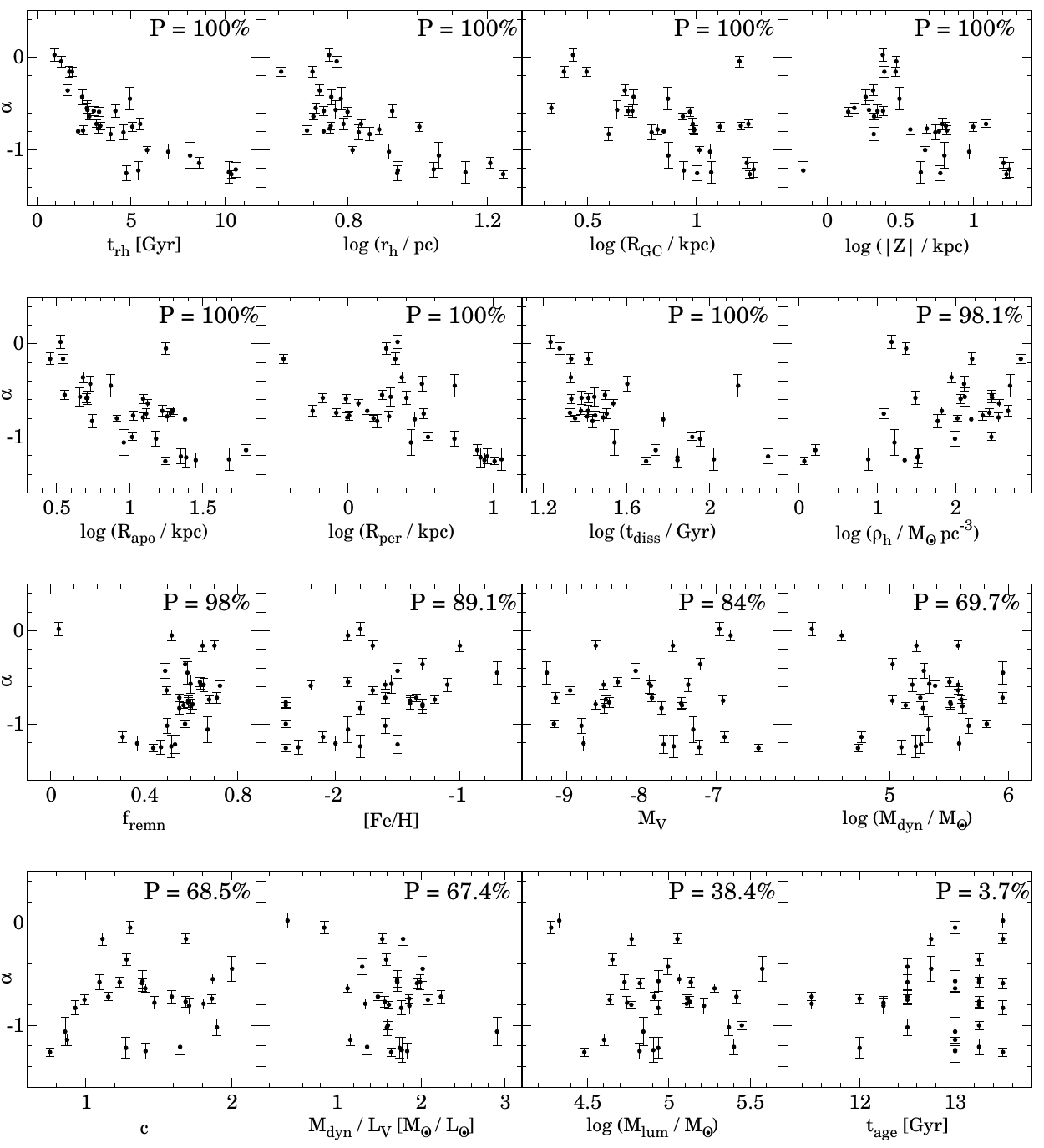}
\caption{Univariate correlations between the global MF slope $\alpha$ and various dynamical, structural and orbital parameters. The statistical significance $(P)$ of each correlation is indicated.}
\label{correlation1}
\end{center}
\end{figure*}

Our sample of GCs constitutes a large
enough data set to investigate the correlations
between MF slope and the dynamical, structural and orbital parameters.
The following parameters have been considered:
the Galactocentric distance $(R_{\text{GC}})$, distance from Galactic plane $(Z)$, 
the average apogalactic ($R_{\text{apo}}$) and perigalactic ($R_{\text{per}}$) distance and
the average dissolution time ($t_{\text{diss}}$) derived by \cite{baumgardt2019}, 
the concentration $(c)$ derived from 
\cite{mcLaughlin2005}, the age $(t_{\text{age}})$ and metallicity ([Fe/H]) derived
by \cite{dotter2010}, the luminous mass ($M_{\text{lum}}$), dynamical mass ($M_{\text{lum}}$),
remnant mass fraction ($f_{\text{remn}}$), V-band magnitude ($M_V$), mass-to light ratio 
($M_{\text{dyn}}/L_V$), half-mass radius ($r_h$), half-mass relaxation time ($t_{rh}$) and
half-mass density ($\rho_h \equiv 3M_{\text{dyn}}/(8\pi r_h^{3})$), all derived from our
adopted best-fitting model.

We have employed a permutation test to calculate the significance of the 
univariate correlations between MF slope and the other parameters. 
For each parameter, an error-weighted least-squares fit has been performed 
and the corresponding $\chi^{2}$ value has been estimated. The same process
has been performed on $10^{4}$ realizations simulated by randomly 
swapping the values of the independent variable. The significance of the 
correlation is defined by the fraction of realizations with a $\chi^{2}$ 
larger than the one calculated in the observed sample. 
The complete set of correlations and their related probabilities 
are shown in Figs \ref{correlation1}.
We found a tight anticorrelation with $P>99.7 \%$ between the present-day MF slope
and the following parameters: $t_{rh}$, $r_h$, $R_{\text{GC}}$, $\log \vert Z \vert$, $\log R_{\text{apo}}$, 
$\log R_{\text{per}}$ and $\log t_{\text{diss}}$, and marginal ($95\% < P < 99.7\%$)
correlations with $\log \rho_h$ and $f_{\text{remn}}$.  
Note that most of these parameters are correlated to each other so that it is not 
easy to distinguish the leading parameter determining the correlation.

\begin{figure*}
\begin{center}
\includegraphics[width=180mm]{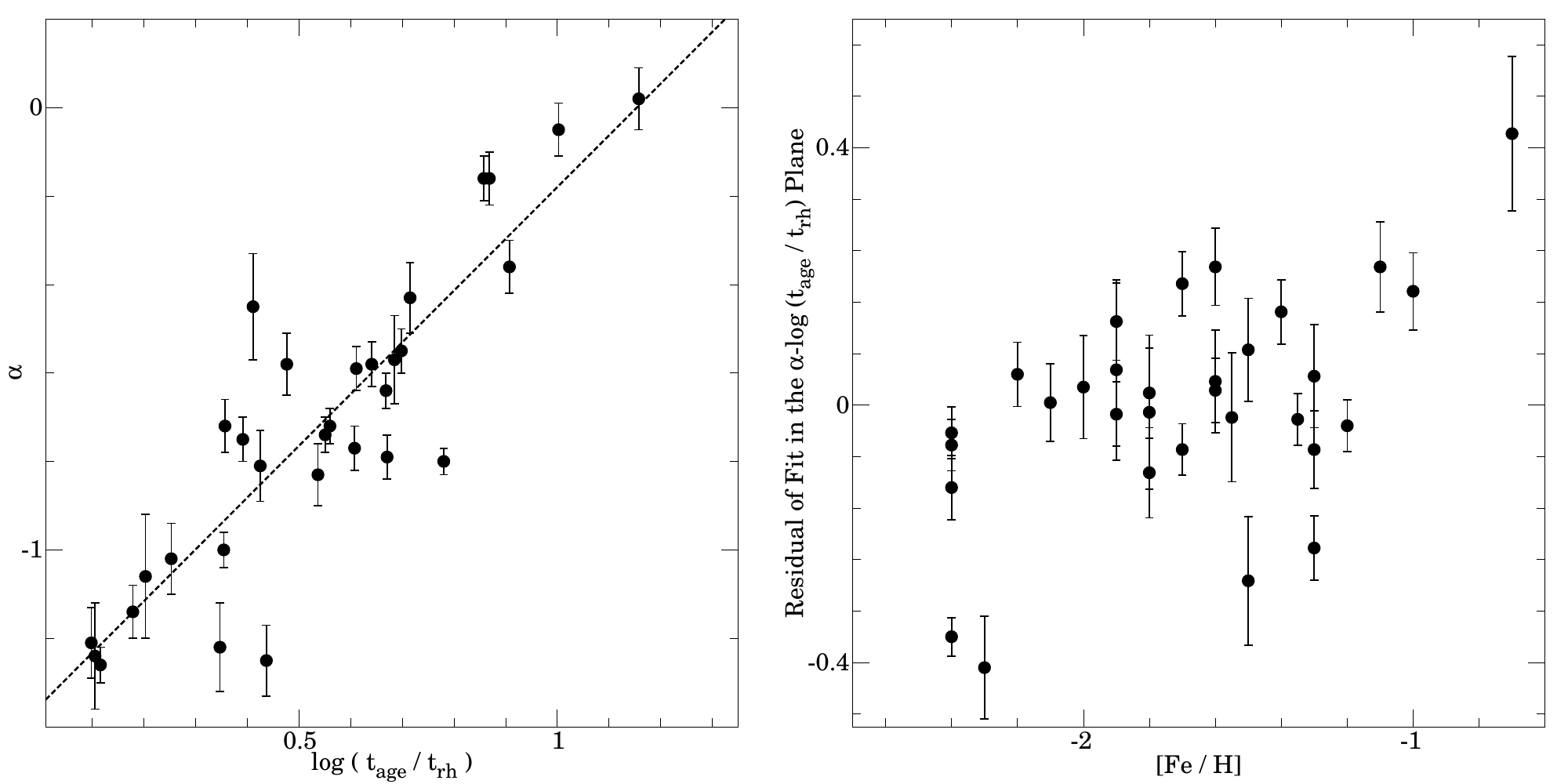}
\caption{Left panel: Slope of GC MFs as a function of the ratio of lifetime to present-day 
half-mass relaxation time. The least-square fit is marked by the dashed line. 
Right panel: Residual of fit in the $\alpha-\log(t_{\text{age}}/t_{rh})$ plane 
against metallicity.} 
\label{corr_res}
\end{center}
\end{figure*}

The anticorrelation with half-mass relaxation
time confirms what was found in \citeauthor{sollima2017b}. The
left panel of Fig. \ref{corr_res} shows another view of the correlation between MF slope
and the ratio between the GC age and present-day half-mass relaxation time. 
The tight correlation is clear:
less-evolved GCs have the steeper MFs. 
It is apparent in this plot that an increase of 
$\alpha$ by a factor of 2 between $\alpha\approx -1.2$ and $\alpha\approx -0.6$
implies an increase of $t_{\text{age}}/t_{rh}$ by a factor of 5. 
Only a marginal correlation is instead found with the remnant fraction, as 
claimed in \citet{sollima2017b}.

Moreover, the 
correlations with the present-day Galactic position \citep[found by][]{djorgovski1993,
capaccioli1993,piotto1999} are also confirmed.

Instead, the correlation with concentration, 
previously reported by \citet{demarchi2007}, is found to be not significant.

As explained above, there are a lot of significant correlations between 
individual parameters and the MF slope. Some of these parameters are 
expressible in terms of each other so that the significance of their univariate 
correlation with the MF slope cannot be interpreted as a proof of the 
physical dependence of the MF on these parameters.
For instance, both the Galactocentric distance 
and half-mass relaxation time
separately show significant correlations with the MF slope. However, GCs at large 
Galactocentric distances
have on average large half-mass radii \citep{vdb1991} and 
thus long relaxation times. So, the correlation with Galactocentric distance can 
be driven by the effect of the relaxation time rather than by the current 
location in the Galaxy.

To avoid this problem, we also considered the bivariate correlations of all 
the possible pairs of parameters listed above.
In this case, for any pair of variables a bilinear fit is performed providing a 
$\chi_{biv}^{2}$. A permutation test by randomly swapping the second independent variable 
is then performed providing $10^{4}$ individual $\chi_{perm}^{2}$. 
The second independent variable is considered an independent correlator if the 
fraction of random realizations with a $\chi_{perm}^{2}>\chi_{biv}^{2}$ 
is larger than 99.7\%.
This test allows to identify the strongest among two correlating 
variables.
Indeed, if we adopt $t_{\text{age}}/t_{rh}$ as the first independent variable,
the only significant bivariate correlator would be metallicity. 
Instead, if we assumed any other parameter as the first
independent variable, we would find the only significant bivariate
correlation using $t_{rh}$ as the second independent variable.
This indicates that the driving parameter in determining the MF slope is 
the half-mass relaxation time, while all the other correlations are only 
consequences of implicit covariances.
In other word, the residuals of a fit in the $\alpha-t_{rh}$ plane 
correlate with metallicity only. 

In the right panel of Fig \ref{corr_res}, the residual of the least-square fit in
the $\alpha-\log(t_{\text{age}}/t_{rh})$ plane 
\begin{equation*}
\alpha=(1.17\pm 0.12)\log (t_{age}/t_{rh})-(1.35\pm 0.07)
\end{equation*}
are shown as a function of 
metallicity. The slope of the apparent trend is $\text{d}\alpha/\text{d[Fe/H]}\sim 0.37$.
We note that most of this correlation is driven by two GCs at the 
metal-poor end (NGC 4590 and NGC 7099) and by the most metal-rich GC of 
our sample (NGC 104).

\section{Discussion}\label{Sec:6}

In this paper, we performed the deepest photometry of HST/ACS/WFC parallel fields
for 32 Galactic GCs, combined them with HST data available for the central 
regions of 
these GCs from the \emph{ACS Treasury Program} \citep{sarajedini2007}, and compared this dataset with multimass dynamical models to
derive the present-day global MFs. Additionally, the masses, mass-to-light ratios, 
half-mass radii and fraction of remnants have been estimated
for 31 GCs with available radial velocity information.

This represents one of the largest data sets for the GC present-day
MFs constraining its variation with measurements in the outer regions of GCs. 
The MF slopes within our sample vary 
in the range $-1.2 \lesssim \alpha \lesssim 0$ and are comparable with the MFs 
found by \citeauthor{sollima2017b}, who estimated MFs by focusing
only on the GCs central region. 
This work represents an improvement with respect to the work by 
\citeauthor{sollima2017b} since most GCs of their sample extend far outside the field 
of view of the ACS, possibly under/over-estimating the degree of mass segregation \citep[see ][]{sollima2015}. 
However, our result showed
that the average difference between the two works is  
$\Delta \alpha = 0.03$ which is even smaller than the formal errors  
of the MF slope for each GC. Nevertheless, we have estimated flatter 
MFs for a few GCs (actually the most extended of the \citeauthor{sollima2017b} sample).

The estimated masses and half-mass radii are in good agreement with those 
found by \citeauthor{sollima2017b} and \citeauthor{baumgardt2018}, although significant 
differences are apparent for some extended GCs.

By analyzing the radial variation of the MF slopes
for each GC, we found that King-Michie multimass models adequately
reproduce the observations for most GCs, but  
they underestimate the degree of mass segregation in dense GCs. 
While we cannot exclude that this evidence is an artifact due to an over-estimation of the 
completeness in the crowded central regions of denser GCs, this evidence could 
be real and indicate that GCs with denser cores evolve faster and reach a larger
degree of mass segregation than what is predicted by 
King-Michie models. In this respect, a similar result has been found by 
\citet{sollima2017a} comparing various snapshots of N-body simulations selected at 
different stages of evolution with the same family of multimass models adopted 
here and found that over a long time interval, as the GC approaches the core-collapse phase, 
simulations reach a degree of mass segregation much larger than that indicated by the best fit 
analytical models.

We evaluated the present-day Jacobi radii for our GCs 
and estimated that, while around 60\%
of them are tidally underfilling with $r_h/r_J\lesssim 0.08$, 
in the outer Galactic region (at $R_{GC} > 8 \,\text{kpc}$) two distinct groups are 
distinguishable:
the tidally underfilling and compact GCs with mean half-mass density   
$\rho_{h,m} \approx 110\, \text{M}_{\odot} \text{pc}^{-3}$ and 
the tidally filling and extended GCs with 
$\rho_{h,m} \approx 6\, \text{M}_{\odot} \text{pc}^{-3}$.  
The GCs in the second group have small masses and are closer to dissolution.
The existence of these two groups of GCs was put forward by 
\citet{baumgardt2010} on the basis of a sample of half-mass and Jacobi radii 
calculated using single-mass fit to the projected density profile and assuming 
a simplified Galactic potential. They interpreted this evidence as a dichotomy 
in the primordial size of proto-GCs, similarly to what is observed in dwarf 
galaxies and young clusters in the Milky Way \citep{dacosta2009,pfalzner2009}.
Unfortunately, only a few GCs of our sample populate the region at large 
Galactocentric distances, where most of the tidally-filling GCs reside in the 
\citet{baumgardt2010} sample. In particular, only 3 GCs classified by 
\citet{baumgardt2010} as tidally filling (NGC 288, NGC 5053 and NGC 5466) are 
included in our sample, two of them have $r_{h}/r_{J}>0.08$.

By investigating possible correlations between various parameters, we found
that MF slope correlates significantly with half-mass relaxation time as
previously suggested by \cite{paust2010} and \citeauthor{sollima2017b}. 
Other correlations have been found to be less significant and mainly 
driven by implicit correlations with $t_{rh}$ \citep[like the ones with the 
present-day positions of GCs in the Galaxy, $R_{GC}$ and $Z$, claimed 
by ][and those with the average apo-/perigalactic distances]{djorgovski1993, 
capaccioli1993, piotto1999}.
This result confirms the theoretical predictions provided by several studies 
\citep{gieles2011,baumgardt2003}, indicating that the MF slope is controlled by
two-body relaxation and evaporation across the cluster tidal boundary.
In particular, two-body relaxation leads to an efficient exchange of 
kinetic energy between GC stars, with the less massive stars gaining orbital energy.
These stars reach the border of the GC potential well and escape in a time-scale
proportional to the inverse of the squared energy \citep{fukushige2000}.
Hence, the short relaxation time causes
a fast and efficient depletion of low-mass stars in the MF (see left panel of Fig. \ref{corr_res}).

We investigate the significance of bivariate correlations between all the 
possible pairs of parameters. We found that, adopting the relaxation time as a 
first correlator, a significant bivariate correlation is found with metallicity. 
This second-order correlation strongly depends on a few GCs lying at the 
extremes of the metallicity range covered by our sample. Its significance is 
therefore not clear. However, if true, this 
correlation would indicate a flatter MF for more metal-rich GCs, in agreement 
with the predictions of star formation theories on the dependence of the IMF slope on metal content 
\citep{silk1977,marks2012,chabrier2014}.

We do not find any correlation between $\alpha$ 
and concentration, as claimed by \cite{demarchi2007}. Such a correlation 
has not been found in any of the studies based on HST datasets, i.e. 
\cite{paust2010} and \citeauthor{sollima2017b}.

The analysis presented here represents a further step forward in the determination of 
the dynamical properties of GCs, in particular their MF. While improvements 
on the observational side (like a better photometric
coverage of the cluster extent, a larger sample of kinematics information, etc.) 
would be valuable to further reduce uncertainties and the possible 
bias, the next important improvement is likely linked to the theoretical 
modelling of these stellar systems. In particular, the results presented here 
indicate that analytical models provide only a 
first-order approximation of the mass segregation occurring in GCs. This might
lead to significant systematics in the determined 
structural/dynamical 
quantities.
Sophisticated techniques like direct N-body fitting (\citeauthor{baumgardt2018}) 
including a realistic treatment of tidal interaction, the development of velocity 
anisotropy and the effect of binary interactions are 
becoming feasible in the near future and represent the next step-forward in the 
determination of 
structural/dynamical 
parameters of GCs.


\section*{Acknowledgments}

We warmly thank the anonymous referee for his/her helpful comments and suggestions.
HE thanks the Iran's Ministry of Science, Research and Technology for providing fellowship in support 
of this research and gratefully acknowledge support by the INAF Osservatorio di Astrofisica e Scienza dello Spazio di Bologna.
We also thank Paolo Montegriffo and Barbara Lanzoni for their helpful support.


\bsp \label{lastpage} 
\begin{thebibliography}{99}

\bibitem[\protect\citeauthoryear{Aguilar, Hut, \& Ostriker}{1988}]{aguilar1988} 
Aguilar L., Hut P., Ostriker J.~P., 1988, ApJ, 335, 720 

\bibitem[\protect\citeauthoryear{Alexander et al.}{2014}]{alexander2014}
Alexander P.~E.~R., Gieles M., Lamers H.~J.~G.~L.~M., Baumgardt H., 2014, MNRAS, 442, 1265

\bibitem[\protect\citeauthoryear{Anderson et al.}{2008}]{anderson2008}
Anderson, J., Sarajedini A., Bedin L.~R., King I.~R., et al., 2008, AJ, 135, 2055

\bibitem[\protect\citeauthoryear{Bastian, Covey \& Meyer}{2010}]{bastian2010}
Bastian N., Covey K.~R., Meyer M.~R., 2010, ARA\&A, 48, 339

\bibitem[\protect\citeauthoryear{Baumgardt \& Makino}{2003}]{baumgardt2003}
Baumgardt H., Makino J., 2003, MNRAS, 340, 227

\bibitem[\protect\citeauthoryear{Baumgardt et al.}{2010}]{baumgardt2010}
Baumgardt H., Parmentier G., Gieles M., Vesperini E., 2010, MNRAS, 401, 1832

\bibitem[\protect\citeauthoryear{BH18}{2018}]{baumgardt2018}
Baumgardt H., Hilker M., 2018, MNRAS, 478, 1520 (BH18)

\bibitem[\protect\citeauthoryear{Baumgardt et al.}{2019}]{baumgardt2019}
Baumgardt H., Hilker M., Sollima A., Bellini A., 2019, MNRAS, 482, 5138

\bibitem[\protect\citeauthoryear{Bergbusch \& VandenBerg}{1992}]{bergbusch1992}
Bergbusch P.~A., VandenBerg D.~A., 1992, ApJS, 81, 163

\bibitem[\protect\citeauthoryear{Brent}{1973}]{brent1973}
Brent R.~P., 1973, in "Algorithm for Minimization without Derivatives", Englewood Cliffs, Prentice-Hall NJ, Chapt. 7

\bibitem[\protect\citeauthoryear{Capaccioli, Piotto \& Stiavelli}{1993}]{capaccioli1993}
Capaccioli M., Piotto G., Stiavelli M., 1993, MNRAS, 261, 819

\bibitem[\protect\citeauthoryear{Cappellari et al.}{2006}]{cappellari2006}
Cappellari M., Bacon R., Bureau M., Damen M.~C., et al., 2006, MNRAS, 366, 1126

\bibitem[\protect\citeauthoryear{Chabrier}{2003}]{chabrier2003}
Chabrier G., 2003, PASP, 115, 763

\bibitem[\protect\citeauthoryear{Chabrier, Hennebelle \& Charlot}{2014}]{chabrier2014} 
Chabrier G., Hennebelle P., Charlot S., 2014, ApJ, 796, 75

\bibitem[\protect\citeauthoryear{Conroy, van Dokkum \& Villaume}{2017}]{conroy2017} 
Conroy C., van Dokkum P.~G., Villaume A., 2017, ApJ, 837, 166 

\bibitem[\protect\citeauthoryear{Czekaj et al.}{2014}]{czekaj2014} 
Czekaj M.~A., Robin A.~C., Figueras F., Luri X., Haywood M., 2014, A\&A, 564, A102 

\bibitem[\protect\citeauthoryear{Da Costa et al.}{2009}]{dacosta2009} 
Da Costa G.~S., Grebel E.~K., Jerjen H., Rejkuba M., Sharina M.~E., 2009, AJ, 137, 4361 

\bibitem[\protect\citeauthoryear{Da Rio et al.}{2012}]{dario2012} 
Da Rio N., Robberto M., Hillenbrand L.~A., Henning T., Stassun K.~G., 2012, ApJ, 748, 14

\bibitem[\protect\citeauthoryear{de Marchi \& Paresce}{1995}]{demarchi1995} 
de Marchi G., Paresce F., 1995, A\&A, 304, 211 

\bibitem[\protect\citeauthoryear{de Marchi, Paresce \& Portegies Zwart}{2005}]{demarchi2005}
de Marchi G., Paresce F., Portegies Zwart S., 2005, in Corbelli E., Palle F., Zinnecker H., eds, 
Astrophysics and Space Science Library, Vol. 327, The Initial Mass Function 50 
years Later, Dordrecht, Springer, p 77

\bibitem[\protect\citeauthoryear{de Marchi, Paresce, \& Pulone}{2007}]{demarchi2007} 
de Marchi G., Paresce F., Pulone L., 2007, ApJ, 656, L65 

\bibitem[\protect\citeauthoryear{Djorgovski, Piotto, \& Capaccioli}{1993}]{djorgovski1993} 
Djorgovski S., Piotto G., Capaccioli M., 1993, AJ, 105, 2148 

\bibitem[\protect\citeauthoryear{Dolphin}{2000}]{dolphin2000}
Dolphin A.~E., 2000, PASP, 112, 1383

\bibitem[\protect\citeauthoryear{Dolphin}{2016}]{dolphin2016}
Dolphin, A. 2016, DOLPHOT: Stellar photometry, Astrophysics Source Code Library, ascl:1608.013

\bibitem[\protect\citeauthoryear{Dotter et al.}{2007}]{dotter2007}
Dotter A., Chaboyer B., Jevremovi\'{c} D., Baron E., et al., 2007, AJ, 134, 376

\bibitem[\protect\citeauthoryear{Dotter et al.}{2010}]{dotter2010}
Dotter A., Sarajedini A., Anderson J., Aparicio A., et al., 2010, ApJ, 708, 698

\bibitem[\protect\citeauthoryear{Frank, Grebel \& K\"{u}pper}{2014}]{frank2014}
Frank M.~J., Grebel E.~K., K\"{u}pper A.~H.~W., 2014, MNRAS, 443, 815

\bibitem[\protect\citeauthoryear{Fukushige \& Heggie}{2000}]{fukushige2000}
 Fukushige T., Heggie D.~C., 2000, MNRAS, 318, 753 

\bibitem[\protect\citeauthoryear{Gennaro et al.}{2018}]{gennaro2018} 
Gennaro M., et al., 2018, ApJ, 855, 20 

\bibitem[\protect\citeauthoryear{Gieles, Heggie \& Zhao}{2011}]{gieles2011}
Gieles M., Heggie D.~C., Zhao H., 2011, MNRAS, 413, 2509

\bibitem[\protect\citeauthoryear{Giersz \& Heggie}{1996}]{giersz1996}
Giersz M., Heggie D.~C., 1996, MNRAS, 279, 1037

\bibitem[\protect\citeauthoryear{Giersz}{2001}]{giersz2001} 
Giersz M., 2001, MNRAS, 324, 218 

\bibitem[\protect\citeauthoryear{Gunn \& Griffin}{1979}]{gunn1979}
Gunn J.~E., Griffin R.~F., 1979, AJ, 84, 752

\bibitem[\protect\citeauthoryear{Haghi et al.}{2014}]{haghi2014}
Haghi H., Hoseini-Rad S.~M., Zonoozi A.~H., K\"{u}pper A.~H.~W., 2014, MNRAS, 444, 3699

\bibitem[\protect\citeauthoryear{Haghi et al.}{2015}]{haghi2015}
Haghi H., Zonoozi A.~H., Kroupa P., Sambaran B., Baumgardt H., 2015, MNRAS, 454, 3872 

\bibitem[\protect\citeauthoryear{Harris}{1996}]{harris1996}
Harris W.~E., 1996, AJ, 112, 1487

\bibitem[\protect\citeauthoryear{Johnston, Spergel \& Hernquist}{1995}]{johnston1995}
Johnston K.~V., Spergel N.~S., Hernquist L., 1995, ApJ, 451, 598

\bibitem[\protect\citeauthoryear{Joshi, Nave \& Rasio}{2001}]{joshi2001}
Joshi K.~J., Nave C.~P., Rasio F.~A., 2001, ApJ, 550, 691

\bibitem[\protect\citeauthoryear{Kaliari et al.}{2009}]{kaliari2009}
Kaliari J.~S., Saul Davis D., Richer H.~B., Bergeron ., et al., 2009, APJ, 705, 408

\bibitem[\protect\citeauthoryear{King}{1966}]{king1966}
King I.~R., 1966, AJ, 71, 64

\bibitem[\protect\citeauthoryear{Kroupa, Tout \& Gilmore}{1993}]{kroupa1993}
Kroupa P., Tout C.~A., Gilmore G., 1993, MNRAS, 262, 545

\bibitem[\protect\citeauthoryear{Kroupa}{2001}]{kroupa2001}
Kroupa P., 2001, MNRAS, 322, 231

\bibitem[\protect\citeauthoryear{Kroupa}{2002}]{kroupa2002}
Kroupa P., 2002, Science, 295, 82 

\bibitem[\protect\citeauthoryear{Kroupa et al.}{2013}]{kroupa2013}
Kroupa P., Weidner C., Pflamm-Altenburg J., Thies I., et al., 2013, Planets, Stars and Stellar Systems:
Volume 5. Springer, Dordrecht, p. 115

\bibitem[\protect\citeauthoryear{Lamers, Baumgardt \& Gieles}{2013}]{lamers2013}
Lamers H.~J.~G.~L.~M., Baumgardt H., Gieles M., 2013, MNRAS, 433, 1378

\bibitem[\protect\citeauthoryear{Madrid, Hurley, \& Martig}{2014}]{madrid2014} 
Madrid J.~P., Hurley J.~R., Martig M., 2014, ApJ, 784, 95 

\bibitem[\protect\citeauthoryear{Mar{\'{\i}}n-Franch et al.}{2009}]{marinfranch2009} 
Mar{\'{\i}}n-Franch A., et al., 2009, ApJ, 694, 1498 

\bibitem[\protect\citeauthoryear{Marks et al.}{2012}]{marks2012} 
Marks M., Kroupa P., Dabringhausen J., Pawlowski M.~S., 2012, MNRAS, 422, 2246 

\bibitem[\protect\citeauthoryear{Massey}{2003}]{massey2003}
Massey P., 2003, ARA\&A, 41, 15

\bibitem[\protect\citeauthoryear{McLaughlin \& van der Marel}{2005}]{mcLaughlin2005}
McLaughlin D.~E., van der Marel R.~P., ApJS, 161, 304


\bibitem[\protect\citeauthoryear{Michie}{1963a}]{michie1963}
Michie R.~W., 1963a, MNRAS, 125, 127

\bibitem[\protect\citeauthoryear{Milone, et al.}{2012}]{milone2012} Milone A.~P., et al., 2012, A\&A, 540, A16

\bibitem[\protect\citeauthoryear{Milone et al.}{2017}]{milone2017}
Milone A.~P., Piotto G., Renzini A., Marino A.~F., et al., 2017, MNRAS, 464, 3636

\bibitem[\protect\citeauthoryear{Murphy, Cohn \& Lugger}{2011}]{murphy2011}
Murphy B.~W., Cohn H.~N., Lugger P.~M., 2011, ApJ, 732, 67

\bibitem[\protect\citeauthoryear{Miller \& Scalo}{1979}]{miller1979}
Miller G.~E., Scalo J.~M., 1979, ApJS, 41, 513

\bibitem[\protect\citeauthoryear{Mor et al.}{2019}]{mor2019} 
Mor R., Robin A.~C., Figueras F., Roca-F{\`a}brega S., Luri X., 2019, A\&A, 624, L1 

\bibitem[\protect\citeauthoryear{Moraux et al.}{2003}]{moraux2003}
Moraux E., Bouvier J., Stauffer J.~R., Cuillandre J.~-C., 2003, A\&A, 400, 891

\bibitem[\protect\citeauthoryear{Nardiello et al.}{2018}]{nardiello2018}
Nardiello D., Libralato M., Piotto G., Anderson J., et al., 2018, MNRAS, 481, 3382

\bibitem[\protect\citeauthoryear{Ostriker, Spitzer, \& Chevalier}{1972}]{ostriker1972} 
Ostriker J.~P., Spitzer L., Jr., Chevalier R.~A., 1972, ApJ, 176, L51 

\bibitem[\protect\citeauthoryear{Paust et al.}{2010}]{paust2010}
Paust N.~E.~Q., Reid I.~N., Piotto G., Aparicio A., et al., 2010, ApJ, 139, 476

\bibitem[\protect\citeauthoryear{Pfalzner}{2009}]{pfalzner2009} 
Pfalzner S., 2009, A\&A, 498, L37 

\bibitem[\protect\citeauthoryear{Piotto \& Zoccali}{1999}]{piotto1999} 
Piotto G., Zoccali M., 1999, A\&A, 345, 485 

\bibitem[\protect\citeauthoryear{Piotto et al.}{2015}]{piotto2015}
Piotto G., Milone A.~P., Bedin L. R., Anderson J., et al., 2015, AJ, 149, 91

\bibitem[\protect\citeauthoryear{Robin et al.}{2003}]{robin2003}
Robin A.~C., Reyl\'{e} C., Derri\'{e} S., Picaud S., 2003, A\& A, 409, 523

\bibitem[\protect\citeauthoryear{Rybizki \& Just}{2015}]{rybizki2015} Rybizki J., Just A., 2015, MNRAS, 447, 3880 

\bibitem[\protect\citeauthoryear{Sarajedini et al.}{2007}]{sarajedini2007}
Sarajedini A., Bedin L.~R., Chaboyer B., Dotter A., et al., 2007, AJ, 133, 1658

\bibitem[\protect\citeauthoryear{Salpeter}{1955}]{salpeter1955}
Salpeter E.~E., 1955, ApJ, 121, 161

\bibitem[\protect\citeauthoryear{Schneider et al.}{2015}]{schneider2015}
Schneider F.~R.~N., Izzard R.~G., Langer N., de Mink S.~E., 2015, ApJ, 805, 20

\bibitem[\protect\citeauthoryear{Sheikhi et al.}{2016}]{sheikhi2016}
Sheikhi N., Hasheminia M., Khalaj P., Haghi H., Zonoozi A.~H., Baumgardt H., 2016, MNRAS, 457, 1028 

\bibitem[\protect\citeauthoryear{Silk}{1977}]{silk1977} 
Silk J., 1977, ApJ, 214, 718 

\bibitem[\protect\citeauthoryear{Sirianni et al.}{2005}]{sirianni2005} 
Sirianni M., et al., 2005, PASP, 117, 1049 

\bibitem[\protect\citeauthoryear{Simioni et al.}{2018}]{simioni2018}
Simioni M., Bedin L.~R., Aparicio A., Piotto G., et al., 2018, MNRAS, 476, 271

\bibitem[\protect\citeauthoryear{Sollima, Bellazzini \& Lee}{2012}]{sollima2012}
Sollima A., Bellazzini M., Lee J.~-W., 2012, ApJ, 755, 156

\bibitem[\protect\citeauthoryear{Sollima \& Mastrobuono Battisti}{2014}]{sollima2014}
Sollima A., Mastrobuono Battisti A., 2014, MNRAS, 443, 3513

\bibitem[\protect\citeauthoryear{Sollima et al.}{2015}]{sollima2015}
Sollima A., Baumgardt H., Zocchi A., Balbinot E., et al., 2015, MNRAS, 451, 2185

\bibitem[\protect\citeauthoryear{Sollima et al.}{2017}]{sollima2017a}
Sollima A., Dalessandro E., Beccari G., Pallanca C., 2017a, MNRAS, 464, 3871

\bibitem[\protect\citeauthoryear{SB17}{2017}]{sollima2017b}
Sollima A., Baumgardt H., 2017b, MNRAS, 471, 3668 (SB17)

\bibitem[\protect\citeauthoryear{Sollima}{2019}]{sollima2019} 
Sollima A., 2019, MNRAS, 489, 2377

\bibitem[\protect\citeauthoryear{Spitzer}{1987}]{spitzer1987}
Spitzer L.~Jr, 1987, Dynamical Evolution of Globular Clusters. Princeton Univ. Press, Princeton, NJ

\bibitem[\protect\citeauthoryear{Takahashi \& Lee}{2000}]{takahashi2000} 
Takahashi K., Lee H.~M., 2000, MNRAS, 316, 671 

\bibitem[\protect\citeauthoryear{van den Bergh, Morbey \& Pazder}{1991}]{vdb1991} 
van den Bergh S., Morbey C., Pazder J., 1991, ApJ, 375, 594

\bibitem[\protect\citeauthoryear{van Dokkum et al.}{2017}]{vandokkum2017}
van Dokkum P.~G., Conroy C., Villaume A., Brodie J., Romanowsky A.~J., 2017, ApJ, 841, 68

\bibitem[\protect\citeauthoryear{Watkins et al.}{2015}]{watkins2015}
Watkins L.~L., van der Marel R.~P., Bellini A., Anderson J., 2015, ApJ, 803, 29

\bibitem[\protect\citeauthoryear{Webb et al.}{2014}]{webb2014}
Webb J.~J., Leigh N., Sills A., Harris W.~E., Hurley J.~R., 2014, MNRAS, 442, 1569

\bibitem[\protect\citeauthoryear{Webb \& Leigh}{2015}]{webb2015}
Webb J.~J., Leigh N.~W.~C., 2015, MNRAS, 453, 3278

\bibitem[\protect\citeauthoryear{Webbink}{1985}]{webbink1985} Webbink R.~F., 1985, IAUS, 113, 541 

\bibitem[\protect\citeauthoryear{Weights et al.}{2009}]{weights2009}
Weights D.~J., Lucas P.~W., Roche P.~F., Pinfield D.~J., Riddick, F., 2009, MNRAS, 392, 817

\bibitem[\protect\citeauthoryear{Weisz et al.}{2013}]{weisz2013} 
Weisz D.~R., et al., 2013, ApJ, 762, 123 

\bibitem[\protect\citeauthoryear{Zhang et al.}{2018}]{zhang2018}
Zhang Z.~-Y., Romano D., Ivison R.~J., Papadopoulos P.~P., Matteucci F., 2018, Nature, 558, 260

\bibitem[\protect\citeauthoryear{Zonoozi et al.}{2011}]{zonoozi2011}
Zonoozi A.~H., K\"{u}pper A.~H.~W., Baumgardt H., Haghi H., Kroupa P., Hilker M., 2011, MNRAS, 411, 1989

\bibitem[\protect\citeauthoryear{Zonoozi et al.}{2014}]{zonoozi2014}
Zonoozi A.~H., Haghi H., K\"{u}pper A.~H.~W., Baumgardt H., Frank M.~J., Kroupa P., 2014, MNRAS, 440, 3172

\bibitem[\protect\citeauthoryear{Zonoozi et al.}{2017}]{zonoozi2017}
Zonoozi A.~H., Haghi H., Kroupa P., K\"{u}pper A.~H.~W., Baumgardt H., 2017, MNRAS, 467, 758 

\end{thebibliography}
\end{document}